\documentclass[printer]{aa}
\usepackage{amssymb}
\usepackage{amsmath}
\usepackage{natbib}
\bibpunct{(}{)}{;}{a}{}{,} 
\usepackage{graphicx}
\usepackage{txfonts}
\usepackage{color}
\usepackage{epstopdf}
\usepackage{wasysym}
\usepackage{booktabs}
\usepackage{multirow}
\usepackage{pifont}
\usepackage{color, colortbl}

\definecolor{Gray}{gray}{0.9}
\definecolor{LightCyan}{rgb}{0.88,1,1}

\newcolumntype{a}{>{\columncolor{Gray}}c}
\newcolumntype{b}{>{\columncolor{white}}c}

\begin{document}
\title{Nature of the Galactic centre NIR-excess sources.\\ I. What can we learn from the continuum observations of the DSO/G2 source?}
\author{Michal Zaja\v{c}ek\inst{1,2}, Silke Britzen\inst{2}, Andreas Eckart\inst{1,2}, Banafsheh Shahzamanian\inst{1}, Gerold Busch\inst{1}, Vladimír Karas\inst{3}, Marzieh Parsa\inst{1,2}, Florian Peissker\inst{1}, Michal Dovčiak\inst{3}, Matthias Subroweit\inst{1}, Franti\v{s}ek Dinnbier\inst{1,3}, and Anton Zensus\inst{2}}
\institute{I. Physikalisches Institut der Universit\"at zu K\"oln, Z\"ulpicher Strasse 77, D-50937 K\"oln, Germany \and Max-Planck-Institut f\"ur Radioastronomie (MPIfR), Auf dem H\"ugel 69, D-53121 Bonn, Germany \and Astronomical Institute, Academy of Sciences, Bo\v{c}n\'{\i}~II 1401, CZ-14131~Prague, Czech Republic}

\authorrunning{M.~Zaja\v{c}ek et al.}
\titlerunning{Galactic centre NIR-excess sources. Continuum of the DSO/G2 }
\date{Received 31 January 2017; Accepted 11 April 2017}
\abstract{The Dusty S-cluster Object (DSO/G2) orbiting the supermassive black hole (Sgr~A*) in the Galactic centre has been monitored in both near-infrared continuum and line emission. There has been a dispute about the character and the compactness of the object: interpreting it as either a gas cloud or a dust-enshrouded star. A recent analysis of polarimetry data in $K_{\rm s}$-band ($2.2\,{\rm \mu m}$) allows us to put further constraints on the geometry of the DSO.}{The purpose of this paper is to constrain the nature and the geometry of the DSO.}{We compare 3D radiative transfer models of the DSO with the NIR continuum data including polarimetry. In the analysis, we use basic dust continuum radiative transfer theory implemented in the 3D Monte Carlo code Hyperion. Moreover, we implement analytical results of the two-body problem mechanics and the theory of non-thermal processes.}{We present a composite model of the DSO -- a dust-enshrouded star that consists of a stellar source, dusty, optically thick envelope, bipolar cavities, and a bow shock. This scheme can match the NIR total as well as polarized properties of the observed spectral energy distribution (SED). The SED may be also explained in theory by a young pulsar wind nebula that typically exhibits a large linear polarization degree due to magnetospheric synchrotron emission.}{The analysis of NIR polarimetry data combined with the radiative transfer modelling shows that the DSO is a peculiar source of compact nature in the S cluster $(r \lesssim 0.04\,{\rm pc})$. It is most probably a young stellar object embedded in a non-spherical dusty envelope, whose components include optically thick dusty envelope, bipolar cavities, and a bow shock. Alternatively, the continuum emission could be of a non-thermal origin due to the presence of a young neutron star and its wind nebula. Although there has been so far no detection of X-ray and radio counterparts of the DSO, the analysis of the neutron star model shows that young, energetic neutron stars similar to the Crab pulsar could in principle be detected in the S cluster with current NIR facilities and they appear as apparent reddened, near-infrared-excess sources. The searches for pulsars in the NIR bands can thus complement standard radio searches, which can put further constraints on the unexplored pulsar population in the Galactic centre. Both thermal and non-thermal models are in accordance with the observed compactness, total as well polarized continuum emission of the DSO.}
\maketitle

\section{Introduction}

Since its discovery in 2012 \citep{Gillessen2012} the near-infrared excess and recombination-line emitting source Dusty S-cluster Object also known as G2 (DSO/G2)\footnote{The name G2 first appeared in \citet{burkert2012} to distinguish the source from the first object of a similar type -- G1 \citep{clenet2004}. The acronym DSO (Dusty S-cluster Object) was introduced by \citet{Eckart2013} to stress the dust emission of the source and the overall NIR excess.} has caught a lot of attention because of its highly eccentric orbit around the supermassive black hole associated with the compact radio source Sgr~A* at the Galactic centre. It has been intensively monitored, especially close to its pericentre passage in the spring of $2014$ \citep{Valencia2015,Pfuhl2015}, when it passed the black hole at the distance of about $160\,\rm{AU}$. No enhanced activity of Sgr~A* has been detected so far in the mm \citep{borkar2016}, radio \citep{bower2015}, and X-ray domains  \citep{mossoux2016}; see however \citet{ponti2015} for the discussion of a possible increase in the bright X-ray flaring rate.

 Despite many monitoring programs and analysis, there has been a dispute about the significance of the detection of tidal stretching of the DSO, which has naturally led to a variety of interpretations. A careful treatment of the background emission by \citet{Valencia2015} revealed the DSO as a compact, single-peak emission-line source at each epoch, both shortly before and after the pericentre passage (see however \citeauthor{Pfuhl2015}, \citeyear{Pfuhl2015}). Moreover, the DSO was detected as a compact continuum source in NIR $L$-band by \citet{Witzel2014}, and as a fainter, stable $K_{\rm s}$-band source \citep{Eckart2013,Shahzaman2016}.

Most of the scenarios that have been proposed so far to explain the DSO and related phenomena may be grouped into the three following categories:
\begin{itemize}
  \item[(i)] \textit{core-less cloud/streamer} \citep{Gillessen2012,burkert2012,schartmann2012,Pfuhl2015,Schartmann2015}
  \item[(ii)] \textit{a dust-enshrouded star} \citep{Murray-Clay2012,Eckart2013,Scoville2013,Ballone2013,Zajacek2014,de-colle2014,Valencia2015,Ballone2016}
  \item[(iii)] \textit{binary/binary dynamics} \citep{Zajacek2014,Prodan2015,Witzel2014}
\end{itemize}

The scenarios (i), (ii), (iii), and a few more are summarized in Table~\ref{tab_scenarios} with corresponding references. 

\begin{table*}[tbh]
\Large
\centering
\caption{Overview of proposed scenarios concerning the nature and the formation of the DSO/G2 and a few corresponding papers.}
\resizebox{\textwidth}{!}{  
\begin{tabular}{a|b}
\hline
\hline
\rowcolor{LightCyan}
\textbf{Scenario} & \textbf{Papers}\\
\hline
                                                                     & \citet{Murray-Clay2012,Eckart2013,Scoville2013,Ballone2013}\\ 
\multirow{-2}{*}{\textbf{star with dusty envelope/disc and outflow}} &  \citet{Zajacek2014,de-colle2014,Valencia2015,Ballone2016,Shahzaman2016}\\
\hline
\textbf{binary/binary dynamics} & \citet{Zajacek2014,Prodan2015,Witzel2014,2016MNRAS.460.3494S}\\                                                   
\hline                                              
                                                    &  \citet{Gillessen2012,burkert2012,schartmann2012,Shcherbakov2014}\\     
\multirow{-2}{*}{\textbf{core-less cloud/streamer}} & \citet{Pfuhl2015,Schartmann2015,2015MNRAS.449....2M,2016MNRAS.455.2187M, 2017MNRAS.465.2310M}\\                                                  
                                        
\hline
\textbf{tidal disruption} & \citet{2012ApJ...756...86M,2014ApJ...786L..12G}\\
\hline
\textbf{nova outburst} & \citet{meyer2012}\\
\hline 
\textbf{planet/protoplanet} & \citet{2015ApJ...806..197M,2016ApJ...831...61T}\\
\hline
\end{tabular}
}
\label{tab_scenarios}
\end{table*} 

 The apparent variety of studies may be explained by a lack of information about the intrinsic geometry of the DSO, which makes the problem of determining the DSO nature degenerate, i.e. more interpretations of the source SED and line emission are possible. Also, several observational studies denied the detection of $K$-band $(2.2\,{\rm \mu m})$ counterpart of the object \citep{Gillessen2012,Witzel2014}, which led to very few constraints on the SED, making both scenarios -- core-less cloud and dust-enshrouded star -- theoretically possible \citep{Eckart2013}. On the other hand, \citet{Eckart2013} and \citet{eckart2014a} show the $K$-band detection of the DSO in both VLT and Keck data, which together with the overall compactness of the source in both line \citep{Valencia2015} and continuum emission \citep{Witzel2014} strengthened the hypothesis of a dust-enshrouded star/binary. 

New constraints on the intrinsic geometry of the source has recently been obtained by \citet{Shahzaman2016} thanks to the detection of polarized continuum emission in NIR $K_{\rm s}$ band in the polarimetry mode of the NACO imager at the ESO VLT. In \citet{Shahzaman2016} we also obtained an improved $K_{\rm s}$-band identification of the source in median polarimetry images at different epochs 2008-2012 (before the pericentre passage). The main result is that the DSO is an intrinsically polarized source with a significant polarization degree of $\sim 30\%$, i.e. larger than the typical foreground polarization in the Galactic centre region ($\sim 6\%$), with an alternating polarization angle as the source approaches the position of Sgr~A*.
 
 \begin{figure}[tbh]
   \centering
   \includegraphics[width=0.5\textwidth]{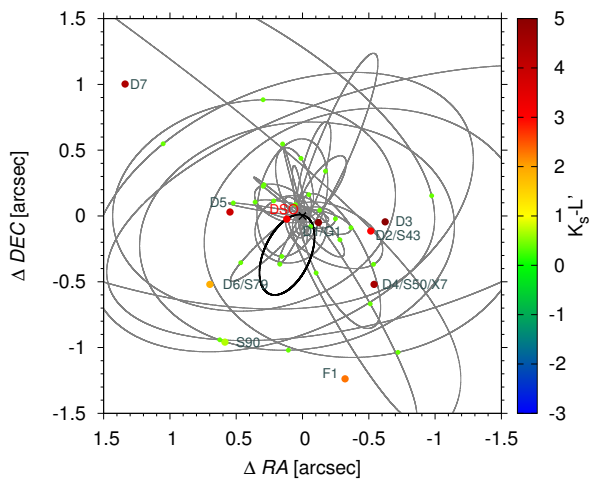}
   \caption{Positions of S stars and infrared excess sources in the innermost $(3.0'' \times 3.0'')$ of the Galactic centre according to \citet{Eckart2013}. The colours denote the colour $(K_{\rm s}-L')$ according to the colour scale to the right. A colour $(K_{\rm{s}}-L')$ for ordinary, B-type S stars (e.g. S2 as a prototype) is expected to be $\sim 0.4$ (with line-of-sight extinction). The position of the DSO and S stars was calculated for $2012.0$ epoch according to the orbital solutions in \citet{Valencia2015} and \citet{Gillessen2009stars}, respectively.}
   \label{fig_nir_excess_positions}
 \end{figure}
 
 Apart from the DSO, \citet{Eckart2013} and \citet{Meyer2014a} showed that the central arcsecond contains several $(\lesssim 10)$ NIR-excess sources, some of which exhibit Br$\gamma$ emission line in their spectra. We show their approximate positions with respect to B-type S stars in Fig.~\ref{fig_nir_excess_positions}. It is not yet clear whether these sources are related to each other, i.e. whether they have a common origin. However, they are definitely peculiar sources with respect to the prevailing population of main-sequence B-type S stars \citep{Eckart1996,Eckart1997,Ghez1998,Gillessen2009stars,2016arXiv161109144G}.
 
 In this paper, we further elaborate on a model of the DSO \citep[see previous models presented in][]{Zajacek2014, zajacek2016} taking into account the new $K_{\rm s}$-band measurements and analysis as presented by \citet{Shahzaman2016}. By comparing theoretical and numerical calculations with the NIR data, we can explain the peculiar characteristics of the DSO by using the model of a young embedded and accreting star surrounded by a non-spherical dusty envelope. This model can be also used for other NIR excess sources, although with a certain caution, since they may be of a different nature (comparison of different formation scenarios is studied in paper II.). 
 
 In this study, we focus on the total as well as linearly polarized NIR continuum characteristics of the DSO. We do not include line radiative transfer in the modelling. We refer the reader to \citet{Valencia2015} and \citet{2015wds..conf...27Z,zajacek2016}, where we studied the basics of the line emission mechanisms potentially responsible for generating broad Br$\gamma$ line. For a pre-main-sequence star, the large line width of the Br$\gamma$ emission line can be explained by the magnetospheric accretion mechanism, where the gas is channelled along the magnetic field lines from the inner parts of a circumstellar accretion disc, reaching nearly free-fall velocities of several $100\,{\rm km\,s^{-1}}$, $v_{\rm infall} \lesssim \sqrt{2GM_{\star}/R_{\star}}$. The line emission can thus be formed within a very compact region of a few stellar radii and the line luminosity is scaled by the accretion luminosity \citep{2014A&A...561A...2A}. 
 
 Furthermore, we also investigate whether the SED of the DSO and that of other excess sources could be of a non-thermal rather than a thermal origin by using the model of a young pulsar wind nebula (PWN). Although the PWN model has a smaller number of parameters and thus a certain elegance in comparison with the model of an embedded star, observationally we miss a clear X-ray or radio counterpart of the DSO, which would be expected for a young neutron star of a few $10^3$ years. On the other hand, our analysis shows that young PWNs resulting from SNII explosions, if present in the nuclear star cluster, could be detected by standard NIR imaging and would indeed manifest themselves as apparent NIR-excess, polarized sources.
 
 The paper is structured as follows. In Section~\ref{sec_constraints} we list important observational characteristics of the DSO. Subsequently, in Section~\ref{sec_compactness}, we briefly analyse the observational as well as theoretical evidence for the compactness of this peculiar source. The results of the modelling and the comparison with observations are presented in Section~\ref{sec_modelling}, where the main focus is on the pre-main-sequence star embedded in a non-spherical dusty envelope (Subsection~\ref{subsec_dust_star}). Moreover, we analyse the possibility that the SED could be of a non-thermal origin, which would open the way for interpreting the DSO as a young pulsar wind nebula (Subsection~\ref{subsec_neutron_star}). In Section~\ref{sec_discussion}, we discuss several other characteristics of the DSO, mainly its association with a larger streamer and other dusty sources as well as further aspects of synchrotron, bremsstrahlung, and Br$\gamma$ luminosity as predicted by a dust-enshrouded star model. Finally, we summarize the main results of the paper I in Section~\ref{sec_summary}.   

\section{Summary of Observational constraints}
\label{sec_constraints}

There are several important constraints that every model of the DSO must explain:
\begin{itemize}
\item[(a)] near-infrared excess or reddening of $K_{\rm s}-L'>3$,
\item[(b)] broad emission lines, especially Br$\gamma$, with the FWHM of the order of $100\,{\rm km\,s^{-1}}$,
\item[(c)] a stable $L'$-band as well as $K_{\rm s}$-band continuum emission,
\item[(d)] a polarized $K_{\rm s}$-band continuum emission of $P_{\rm L} \simeq 30\%$.
\end{itemize}

\begin{figure*}[tbh]
 \centering
 \includegraphics[width=\textwidth]{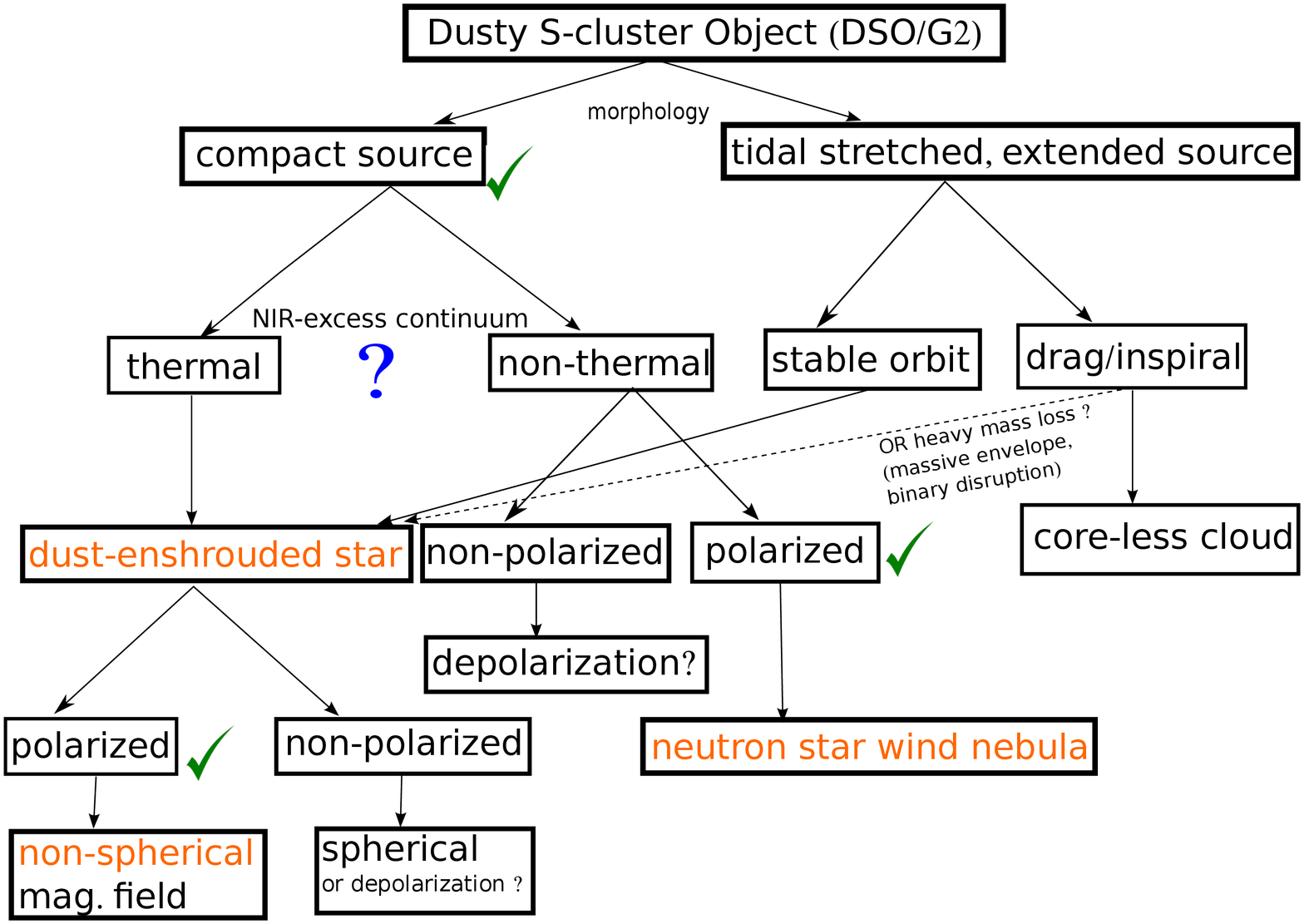}
  \caption{Roadmap for solving the nature of the DSO/G2 source. The ticks (\ding{51}) label the implications of the analysis of observational data by \citet{Valencia2015}, \citet{Witzel2014}, and \citet{Shahzaman2016}. The question-mark ({\bf ?}) on the left implies that the NIR continuum of the DSO can be either \textit{thermal} (standard interpretation) or \textit{non-thermal}. The question-mark ({\bf ?}) on the right stands for potential X-ray and radio counterparts of the DSO that are beyond the detection limit of current instruments. The orange colour marks the possible nature of the DSO that would explain the observed characteristics.}
  \label{fig_roadmap}
\end{figure*}

Besides (a)--(d) characteristics, one should also consider the overall \textit{compactness} or the \textit{diffuseness} of the source (i.e. whether the object can be resolved given the PSF of the instrument used), and the overall orbital evolution (i.e. if one can detect significant drag/inspiral along the orbit as would be expected for a core-less cloud). 

Since different aspects are involved, we set up the roadmap towards solving the DSO nature, which is illustrated in Fig.~\ref{fig_roadmap}.

In the further analysis, we consider the results of \citet{Valencia2015} that show that the DSO exhibits a single-peak $Br\gamma$ emission line at each epoch, i.e. they detect no significant stretching along the orbit as would be expected for a core-less cloud. A consistent result is presented by \citet{Witzel2014}, who detect a compact $L$-band emission of the DSO/G2 during the peribothron\footnote{We use \textit{pericentre} and \textit{peribothron} interchangeably; \textit{peribothron} is derived from Greek \textit{bothros}, which means a hole dug in the ground, also a pit used for drink offerings for subterranean gods in the Greek mythology.} passage in $2014$. 

Therefore, we are not going to consider the tidal stretching, which according to the roadmap in Fig.~\ref{fig_roadmap} would either indicate an extended circumstellar envelope that does not feel the gravitational field of the star, or a core-less cloud.  These two scenarios will be further tested by the orbital evolution in the post-peribothron phase. 

Concerning the spectral energy distribution of the DSO, we adopt the results of \citet{Gillessen2012}, \cite{Eckart2013}, and \citet{Shahzaman2016}; see also \citet{Shcherbakov2014} for the first SED analysis of the DSO assuming a core-less cloud scenario. As analysed by \citet{Eckart2013} and confirmed by \citet{Shahzaman2016}, the source shows prominent reddening between NIR $K_{\rm s}$ and $L'$ bands, $K_{\rm s}-L'>3$. Further measurements in $M$ band were performed by \citet{Gillessen2012}. The $H$-band flux density is an upper limit since there was no detection \citep{Gillessen2012,Eckart2013}.

In the NIR H band, \citet{Eckart2013} derive the minimum apparent magnitude of $m_{\rm H}>21.2$ based on the background level of the neighbouring stars. Using the extinction correction of $A_{\rm H}=4.21$, the upper limit on the flux density is $F_{\rm H} \lesssim 0.17\,{\rm mJy}$. 

For the flux density in NIR $K_{\rm s}$-band, we take a weighed mean of the measurements by \citet{Shahzaman2016} (see our Table 3 for the summary of measurements). We obtain the mean flux density value of $\overline{F}_{2.2}=(0.23\pm 0.02)\,{\rm mJy}$, both with and without considering the epoch 2011, when the ratio $S/N$ was low.

For the $L'$-band magnitude, we consider the value of $m_{\rm L'}=14.4 \pm 0.3$ \citep{Eckart2013}. Using $A_{L'}=1.09$, we obtain the flux density of $F_{3.8}=(1.2\pm 0.3)\,{\rm mJy}$.

The absolute dereddened M-band magnitude obtained by \citet{Gillessen2012} is $M_{M}=-1.8 \pm 0.3$, which yields $F_{4.7}=(1.2 \pm 0.4)\,{\rm mJy}$.

\begin{table}[tbh]
 \centering
 \caption{Summary of NIR flux densities of the DSO. References: 1)~\citet{Eckart2013};  2)~\citet{Eckart2013,Shahzaman2016};  3)~\citet{Gillessen2012,Eckart2013,Witzel2014};  4)~\citet{Gillessen2012}.}
 \resizebox{0.5\textwidth}{!}{  
 \begin{tabular}{ccccc}
   \hline
   \hline
   Band & Wavelength $[{\rm \mu m}]$ & Frequency $[{\rm Hz}]$ & Flux density $[{\rm mJy}]$ & Refs. \\
   \hline
   $H$ & $1.65$ & $1.82\times 10^{14}$  & $\lesssim 0.17$ & 1)  \\       
   $K_{\rm s}$ & $2.2$ & $1.36 \times 10^{14}$ & $0.23 \pm 0.02$ & 2)  \\
   $L'$ & $3.8$ & $7.89 \times 10^{13}$ & $1.2 \pm 0.3$ & 3)  \\
   $M$ & $4.7$ & $6.38 \times 10^{13}$ & $1.2 \pm 0.4$ & 4)  \\
   \hline
 \end{tabular}
 }
 \label{tab_flux_density}
\end{table}

 The overall flux densities are summarized in Table~\ref{tab_flux_density}. Based on the flux density measurements, the upper limit on the overall luminosity was set to be $L_{\rm DSO} <30\,L_{\odot}$ \citep{Eckart2013,Witzel2014}.
 
  Another important constraint is the detection of polarized emission in NIR $K_{\rm s}$ band by \citet{Shahzaman2016} with the polarization degree of $\sim 30\%$ and a variable polarization angle for four consecutive epochs (2008, 2009, 2011, and 2012). The summary of all observational constraints is in Table~\ref{tab_constraints}.

\begin{table}[tbh]
\caption{Summary of observational constraints for the DSO nature.}
\resizebox{0.5\textwidth}{!}{  
  \begin{tabular}{cc}
    \hline
    \hline
    Constraint & Note\\
    \hline
    SED   &  "red" source; $K_{\rm s}-L'>3$\\
    broad emission lines & $FWHM_{\rm Br\gamma}\sim 100\,{\rm km\,s^{-1}}$\\
    source of polarized $K_{\rm s}$ band emission & polarization degree $\sim 30\%$\\
    stability and compactness & no significant tidal elongation\\
    \hline 
  \end{tabular}
  }  
  \label{tab_constraints}
\end{table}

\begin{figure}[tbh]
  \centering
  \includegraphics[width=0.5\textwidth]{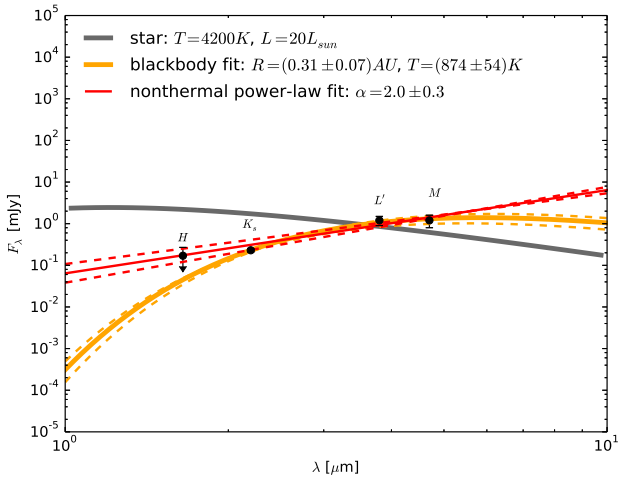}
  \caption{Detected, dereddened flux densities of the DSO (black points) and the fitted continuum: \textit{thermal} black-body fit curve  corresponding to warm dust (orange solid and dashed lines) and \textit{nonthermal} power-law emission fit $S_{\nu} \propto \nu^{-\alpha}$ with the index $\alpha=2.0 \pm 0.3$ (red solid and dashed lines). For comparison, we also plot the black-body curve corresponding to the star of $T_{\star}=4200\,\rm{K}$ and $L_{\star}=20\,L_{\odot}$ (pre-main-sequence star without a dusty envelope; gray solid line).}
  \label{fig_sed_simple}
\end{figure}

The dereddened, continuum flux densities in the NIR domain were attributed to the warm dust emission in the temperature interval $T_{\rm dust}=400\,{\rm K}$--$600\,{\rm K}$ \citep{Gillessen2012,Eckart2013}, which can reproduce the flux densities between $K_{\rm s}$ and $M$ bands. In Fig.~\ref{fig_sed_simple}, we repeat the fit of the dereddened flux densities with a single-temperature black-body flux density profile $S_{\nu}(T,R)=(R/d)^2 \pi B_{\nu}(T)$, where $B_{\nu}(T)$ is a black-body specific intensity at temperature $T$, $R$ is the characteristic radius of the object, and $d$ is the distance to the Galactic centre, which we set to $d=8\,{\rm kpc}$ \citep{Eckart2005, genzel2010, Eckart2017}. The black-body fits gives the temperature of $T=(874 \pm 54)\,{\rm K}$, which corresponds to the warm dust component, and the characteristic radius of $R=(0.31 \pm 0.07)\,{\rm AU}$ for an optically  thick black body. This corresponds to a rather compact source in comparison with the originally proposed mean length-scale of $R_{\rm c}\approx 15\,\rm{mas}\approx 120\,{\rm AU}$ for a core-less gas cloud \citep{Gillessen2012}, in which case the emission would necessarily be optically thin. For comparison, we also plot the black-body curve corresponding to the star of $T_{\star}=4200\,\rm{K}$ and $L_{\star}=20\,L_{\odot}$ (without any circumstellar envelope), which naturally has a reversed slope than that of the DSO continuum (see Fig.~\ref{fig_sed_simple}).  

On the other hand, an alternative model for the origin of the SED with increasing flux values towards longer wavelengths (smaller frequencies) is a nonthermal power law, $S_{\nu}=S_0(\nu/\nu_0)^{-\alpha}$. A naive fit of the power law model to the detected flux densities yields the profile,

\begin{equation}
  S_{\nu} \approx 2.0\, \left(\frac{\nu}{5.37 \times 10^{13}\,{\rm Hz}}\right)^{-2.0\pm 0.3}\,{\rm mJy}\,,
  \label{eq_nonthermal_fit}
\end{equation}   

with partial power-law slopes $\alpha_{ij}=-\log{(S_{\nu i}/S_{\nu j})}/\log{(\nu_i/\nu_j)}$ between bands $i$ and $j$: $\alpha_{HK} \gtrsim 1.05$, $\alpha_{KL}=3.02$, and $\alpha_{LM} \approx 0$. The qualitative behaviour of the continuum radiation of the DSO is similar to the SED of few young pulsars detected in NIR domains \citep{2012A&A...544A.100M,2013A&A...554A.120Z,2016MNRAS.455.1746Z} that are dominated by magnetospheric synchrotron radiation. Moreover, they exhibit a significant linear polarized emission with the polarization degree of a few $10\%$ \citep{2012A&A...542A..12Z}. Thus, DSO-like sources could in principle be neutron stars and the detection of significant linearly polarized emission strengthens this hypothesis. The crucial point for the confirmation of this association is the detection of the counterparts in different domains, specifically radio and X-ray domains.

\section{On the compactness of the DSO}
\label{sec_compactness}

In the literature, it is still debated whether the DSO/G2 source is of compact nature or not. 
In the further analysis we assume that the DSO has a compact nature, i.e. it is not significantly stretched and elongated by tidal forces of the SMBH. This conclusion is based on the fact that both the line emission (Br$\gamma$) and the $L'$-band continuum emission did not show signs of tidal interaction during the closest passage \citep{Valencia2015,Witzel2014}; see however \citet{Pfuhl2015} for a different view. In addition, $K_{\rm s}$-band flux density was constant within uncertainties for four consecutive epochs \citep[2008, 2009, 2011, and 2012][]{Shahzaman2016}.  Since the DSO appears as a point-like source, it is only possible to infer an upper limit for its length-scale based on the minimum FWHM of the point spread function, $\theta_{\rm min} \approx 63 (\lambda/{\rm 2\,\mu m})\,{\rm mas}$.

A simple test of the compactness of the DSO is provided by the comparison of the observed evolution with that of a core-less gaseous stream approaching the black hole of $M_{\bullet}=4\times 10^6\,M_{\odot}$. When the parts of the cloud move independently in the potential of the black hole, the cloud with initial radius $R_{\rm init}$ is stretched along the orbit and compressed in the perpendicular direction.

\begin{figure*}[tbh]
  \centering
  \begin{tabular}{cc}
  \includegraphics[width=0.5\textwidth]{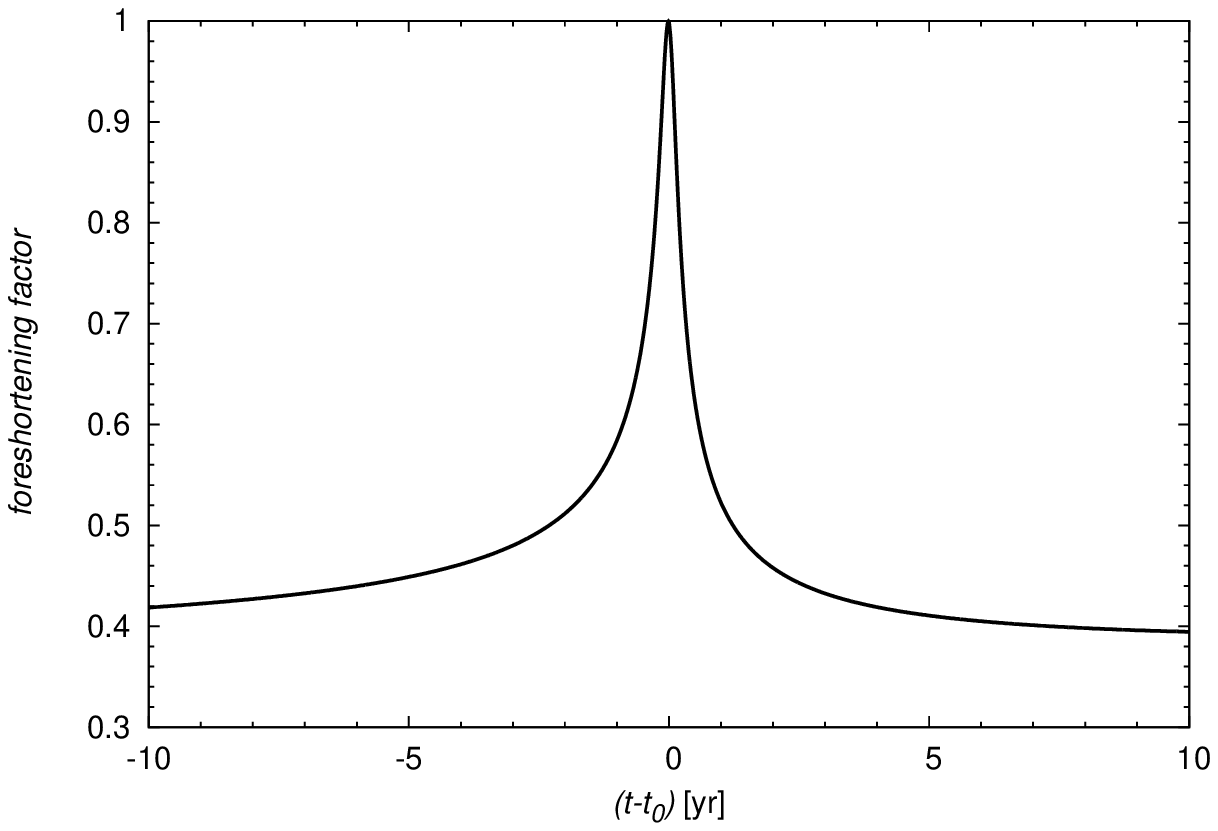} & \includegraphics[width=0.5\textwidth]{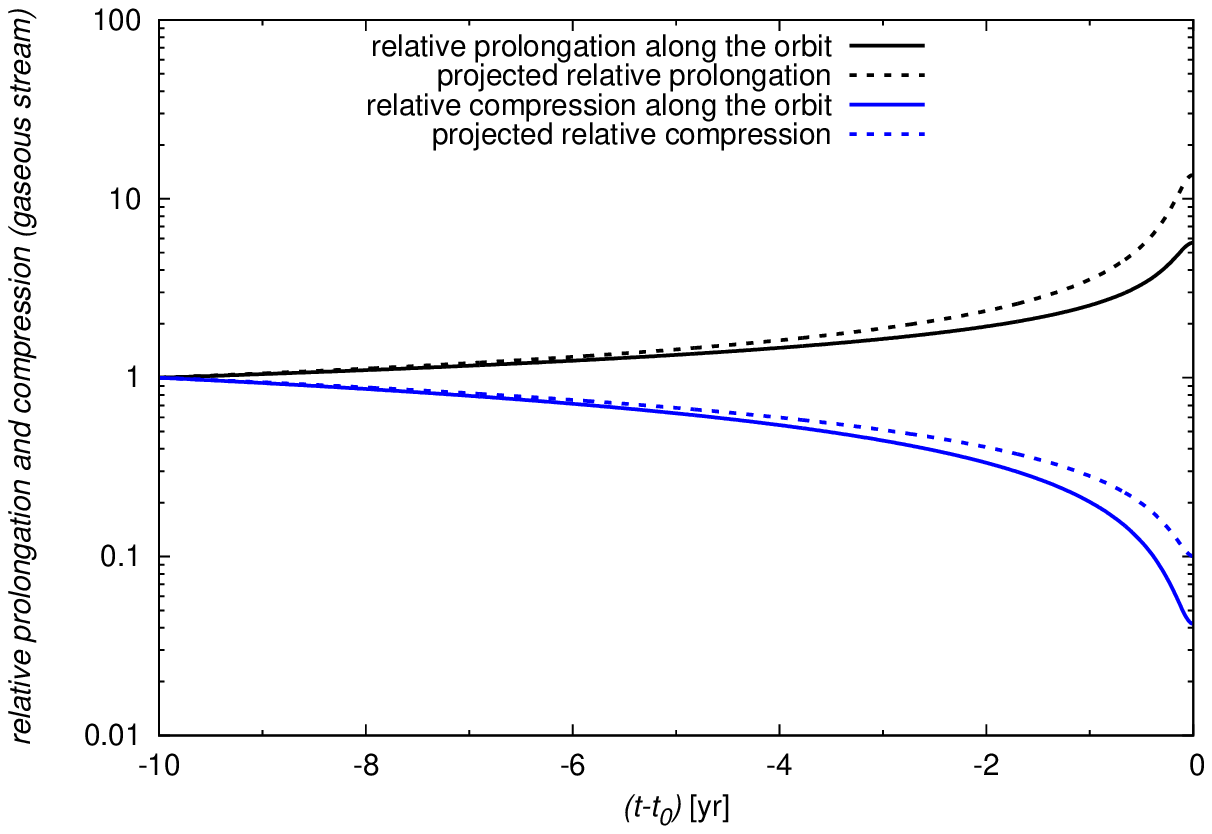}
  \end{tabular}
  \caption{\textbf{Left:} Foreshortening factor for the size of any structure calculated for the DSO orbit with inclination $i=113^{\circ}$. \textbf{Right:} Relative tidal stretching and compression as function of time (in years with respect to the peribothron passage) in the orbital plane (solid lines) and with respect to the inclined orbit to the DSO.}
  \label{fig_tides}
\end{figure*}

According to \citet{Shcherbakov2014}, we can express the relative prolongation as function of distance $r$ from the SMBH in terms of the half-length $L$,

\begin{equation}
  \Lambda=\frac{L}{R_{\rm init}}=\left(\frac{r_{\rm init}}{r}\right)^{1/2}\left(\frac{r_{\rm A}+r_{\rm P}-r}{r_{\rm A}+r_{\rm P}-r_{\rm init}}\right)^{1/2}\,,
  \label{eq_half_length}
\end{equation}
where $r_{\rm init}$ is the initial distance at which the cloud was formed, having a spherical shape with the radius $R_{\rm init}$, and $r_{\rm P}$ and $r_{\rm A}$ are pericentre and apocentre distances of the DSO, respectively. We set $r_{\rm init}$ to the distance that corresponds to 10 years before the peribothron passage (earlier date of formation would lead to larger prolongation). Similarly, the relative compression $\rho$ in the direction perpendicular to the orbit may be expressed as,

\begin{equation}
  \rho=H/R_{\rm init}=\frac{r}{r_{\rm init}}\,.
  \label{eq_compression}
\end{equation}
where $H$ is the actual perpendicular cross-section of the cloud.
For the orbital parameters of the DSO \citep{Valencia2015}, both the relative prolongation $\Lambda(t)$ and compression $\rho(t)$ are plotted in Fig.~\ref{fig_tides} (right panel) as functions of time before the peribothron passage. For the observed dimensions of the object, the foreshortening due to the orbital inclination is important. The foreshortening factor as function of time is plotted in Fig.~\ref{fig_tides} (left panel). At the peribothron, the source should be viewed at full size and the relative prolongation is $\sim 6$ times that of the initial size ($\sim 10$ times when foreshortening is taken into account). The compression in the perpendicular direction should lead to the general spaghettification of the cloud. The tidal elongation of this order was, however, not detected \citep{Witzel2014,Valencia2015}.

 Specifically, between the epoch 2011.39 (3 years before the peribothron) and the peribothron passage, the prolongation factor is $\Lambda \simeq 3.45$. If the DSO was a core-less cloud, it should have a pericentre length-scale  $L_{\rm per}=\Lambda R_{\rm init}$. \citet{Gillessen2013b} infer the FWHM length-scale $R_{\rm FWHM}=(42\pm 10)\,{\rm mas}$ from the Gaussian fit for epochs $2008.0$, $2010.0$, $2011.0$, $2012.0$, $2013.0$. The pericentre size, with respect to 2011 epoch, should then be $L_{\rm per}\approx 145\,{\rm mas}> \theta_{\rm min}$, which is more than a factor of two larger than PSF FWHM. Such a large size is inconsistent with a rather compact line emission detected by \citet{Valencia2015} during the pericentre passage. In fact, the analysis of the $L$-band continuum emission of the DSO during the pericentre passage by \citet{Witzel2014} yields the diameter of $32\, \rm{mas}$, i.e. fully consistent with a point source. 
 
 On the other hand, since the DSO was detected as a point source at the pericentre, the upper limit on its length-scale is given by the diffraction limit $L_{\rm per} \leq \theta_{\rm min}\approx 63\,{\rm mas}$. A core-less cloud or an extended envelope of a star that reached the size of $L_{\rm per}$ by tidal stretching must have been smaller by a factor of $\Lambda(-10\,{\rm yr})\approx 6$, i.e. 10 years before the peribothron passage, yielding the characteristic size of $L(-10\,{\rm yr})\lesssim 10\,{\rm mas}\approx 80\,{\rm AU}$. Based on this estimate, we set the characteristic radius of the potentially tidally perturbed part of the DSO to $R_{\rm c}\lesssim 5\,{\rm mas}$.
 
  Using the observationally inferred Br$\gamma$ luminosity of $L_{\rm Br\gamma}\approx 2\times 10^{-3}\,L_{\odot}$ \citep{Gillessen2012, Valencia2015}, we can estimate the electron density in the envelope assuming case B recombination \citep{Gillessen2012},
\begin{equation}
  n_{\rm e}=1.35 \times 10^6 f_{\rm V}^{-1/2} \left(\frac{R_{\rm c}}{5\,{\rm mas}}\right)^{-3/2} \left(\frac{T_{\rm g}}{10^{4}\,{\rm K}} \right)^{0.54}\,{\rm cm^{-3}}\,,
  \label{eq_electron_density}
\end{equation}  
where $T_{\rm g}$ is the expected gas temperature of the DSO and $f_{\rm V}$ is the volume filling factor, which we set to $f_{\rm V}\leq 1$. For the mass of the DSO, in case it would be a gas cloud, we get
\begin{equation}
  M_{\rm DSO, cloud}\simeq 2.2\times 10^{27} f_{\rm V}^{1/2} \left(\frac{R_{\rm c}}{5\,{\rm mas}}\right)^{3/2} \left(\frac{T_{\rm g}}{10^{4}\,{\rm K}} \right)^{0.54}\,{\rm g}\,
  \label{eq_mass_cloud}
\end{equation}
which is about $M_{\rm DSO,cloud}=0.4f_{\rm V}^{1/2}\,M_{\rm Earth}$. A smaller mass and a required higher density than originally estimated \citep{Gillessen2012} shorten typical hydrodynamical time-scales that determine the lifetime of the cloud, specifically the cloud evaporation time-scale is \citep{burkert2012}

\begin{equation}
  \tau_{\rm evap}=43 \left(\frac{r}{5.4 \times 10^{16}}\right)^{1/6} \left(\frac{M_{\rm DSO,cloud}}{2.2\times 10^{27}\,{\rm g}}\right)^{1/3}\,{\rm yr}\,,
\end{equation}
where $r$ is the distance from Sgr~A* in units of $5.4 \times 10^{16}\,{\rm cm}$, which corresponds approx. to the epoch of $2004$. Such a short evaporation time-scale for a small, cold clump in the hot ambient plasma would necessarily lead to observable changes in the cloud line and continuum luminosities. However, the observations imply that the DSO is a rather compact, stable source in both line and continuum emission \citep{Witzel2014,Valencia2015,Shahzaman2016}.

Although a magnetically arrested cloud was suggested to explain the apparent stability \citep{Shcherbakov2014}, it would still not prevent the cloud from the progressive evaporation and disruption \citep{2015MNRAS.449....2M} as well as the loss of angular momentum when interacting with the ambient medium, leading to the inspiral and deviation from the original ellipse \citep{Pfuhl2015,2015MNRAS.449....2M}.

 Therefore, given the reasons above, a stellar nature of the DSO is more consistent with its observed line and continuum characteristics.


The basic estimate of the distance $r_{t}$ where an object with a characteristic radius $R_{\rm DSO}$ and mass $M_{\rm DSO}$ is tidally disrupted is given by $r_{t}=R_{\rm DSO}(3M_{\bullet}/M_{\rm DSO})^{1/3}$. For a stellar source, we get an estimate $r_{t} \simeq 1.07 (R_{\rm DSO}/1R_{\odot})(M_{\rm DSO}/1M_{\odot})^{-1/3}\,{\rm AU}$. Since the peribothron distance of the DSO is $r_{P}=a(1-e)=0.033\,{\rm pc} \times (1-0.976)\approx 163\,{\rm AU}$ \citep{Valencia2015} and no visible tidal interaction was observed, the upper limit on the length-scale of the stellar system that stays stable is $R_{\rm DSO} \approx 0.7\,(M_{\rm DSO}/1\,M_{\odot})^{1/3}\,{\rm AU}$. On the other hand, for the cloud of $R_{\rm DSO}=15\,\rm{mas} \approx 2.7\times 10^{4}\,R_{\odot}$ and the mass of three Earth mass, $M_{\rm DSO}\approx 10^{-5}\,M_{\odot}$ \citep{Gillessen2012}, the tidal radius is $r_{t}\approx 1.3\times 10^6\,{\rm AU}$. Hence, the cloud should be strongly perturbed not only at the pericentre, but during the whole phase of infall, since the apocentre distance is $r_{\rm A}=a(1+e)\approx 1.3\times 10^4\,\rm{AU}$. In summary, to explain the compact behaviour of the object, the emitting material (gas+dust) should be located in the inner astronomical unit from the star.

On the other hand, since the NIR-continuum is dominated by thermal dust emission (for a dust-enshrouded star), one can estimate the lower distance limit where the dust is located from the dust sublimation radius $r_{\rm sub}$ \citep{2002ApJ...579..694M},

\begin{equation}
  r_{\rm sub}=1.1 \sqrt{Q_{\rm R}}\left(\frac{L_{\rm DSO}}{1000\,L_{\odot}}\right)^{1/2} \left(\frac{T_{\rm sub}}{1500\,{\rm K}}\right)^{-2}\,{\rm AU}\,,
 \label{eq_sublimation}
\end{equation}
where $Q_{\rm R}$ is the ratio of absorption efficiencies, which we consider to be of the order of unity, and $T_{\rm sub}$ is the dust sublimation temperature, for which we take $1500\,{\rm K}$. The inferred luminosity of the DSO is of the order of $L_{\rm DSO}\approx 10\,L_{\odot}$ \citep{Eckart2013,Witzel2014}, and so the expected sublimation radius is $r_{\rm sub}\approx  0.1\,{\rm AU}$. Hence, since the continuum emission seems to be compact and no clear evidence of tidal interaction was detected  during the peribothron passage \citep{Witzel2014}, the distance range, where the dust emitting the NIR-continuum is located and stays gravitationally unperturbed, is quite narrow, $r_{\rm sub}\lesssim r \lesssim r_{\rm t}$, i.e. $0.1\,{\rm AU}\lesssim r \lesssim 1\,{\rm AU}$. 

\section{Modelling the total and polarized continuum emission of the DSO}
\label{sec_modelling}

In this section, we focus on the modelling of the total as well as polarized flux density in corresponding NIR bands. 

The main aim is the continuum radiative transfer in the dense dusty envelope surrounding a star with the emphasis on the \textit{linearly polarized emission} (Section~\ref{subsec_dust_star}). The continuum profile is shaped by reprocessing the stellar emission by the surrounding dusty envelope. Linear polarization may arise due to (i) photon scattering on spherical dust grains (Mie scattering), (ii) dichroic extinction caused by selective absorption of photons in the medium where non-spherical grains are aligned due to radiation or magnetic field. The overall linear polarization degree for young stars may vary from the fraction of a percent up to a few $10\%$, depending on the geometry as well as the extinction of the dusty envelope \citep[see][for a review]{2015psps.book.....K}. 

In case of a hypothetical non-thermal origin of the DSO continuum (Section~\ref{subsec_neutron_star}, one expects a slope of the flux density $S_{\nu} \propto \nu^{-n}$, where $n>0$. Typically, young neutron stars exhibit a multiwavelength non-thermal continuum, which arises due to the synchrotron process in the magnetosphere of young neutron stars. Another contribution is the thermal emission of the cooling surface. However, for young $(\lesssim 10\,{\rm kyr})$ and middle-aged $(\gtrsim 100\,{\rm kyr})$ neutron stars, the non-thermal component dominates in NIR bands. Since neutron stars typically possess a highly-ordered strong magnetic field, the non-thermal component is expected to be partially linearly polarized. Using the homogeneous magnetic field approximation, the linear polarization degree for the synchrotron emission can be determined as \citep{Rybicki1979},

\begin{equation}
  P_{\rm{L}}\lesssim \frac{n+1}{n+5/3}\,,
  \label{eq_neutronstar_pol_deg}
\end{equation}     
where for $n \approx 0.7$ one gets $P_{\rm{L}}\lesssim 0.72$, whereas the measured value in $K_{\rm{s}}$ band (VLT/ISAAC) is $P^{\rm{avg}}_{\rm{L}}\simeq 0.47$ \citep{2012A&A...542A..12Z} for the surrounding pulsar wind nebula.

\subsection{Thermal origin of the SED: DSO as a dust-enshrouded star}
\label{subsec_dust_star}

To find the model of a dust-enshrouded star that reproduces the observed characteristics (SED, broad emission lines, linear polarization, and overall stability and compactness; see also Table~\ref{tab_constraints} for the summary) we perform a set of 3D dust continuum radiative transfer simulations with different components and morphologies of dusty envelopes. 

For solving the radiative transfer equation, we choose a Monte Carlo technique suitable for arbitrary three-dimensional dusty envelopes \citep{2011ffbh.book..151W}. For all our simulations of a dust-enshrouded star, we used an open-source parallelized code \textit{Hyperion} \citep{2011A&A...536A..79R}, which enables to create SEDs as well as images for a required wavelength range as well as different inclinations of the source. Since in the random walk of photons the scattering process is also included, we obtain a full Stokes vector $(I,Q,U,V)$, which enables us to calculate the linear polarization degree as well as the angle according to standard definitions,

\begin{align}
P_{\rm L} & =  \frac{\sqrt{Q^2+U^2}}{I}\,,\nonumber\\
\phi & =  \frac{1}{2} \arctan{\left(\frac{U}{Q} \right)}\,.\label{eq_poldeg_angle}
\end{align} 

Since the extinction is expected to be high in the innermost parts of the envelope, we make use of a modified random walk (MRW) implemented in the code in the thickest regions.

An important part of the modelling is the generation of the dust distribution, for which we use the code \textit{bhmie} by C.F. Bohren and D. Huffman (improved by B. Draine; \citeauthor{bohren1998absorption}, \citeyear{bohren1998absorption}), which provides solutions to the Mie scattering and absorption of light by spherical dust particles. We generate dust grains with a power-law distribution $n(a)\propto a^{-3.5}$ with the smallest and the largest grain size of $(a_{\rm min},a_{\rm max})=(0.01,10.0)\,{\rm \mu m}$. The gas-to-dust mass ratio in the dusty circumstellar envelope for all geometries is assumed to be 100. The distance to the Galactic centre is set to $8\,\rm{kpc}$ for flux density calculations.

For most of the simulations, we set up a 3D spherical grid that contains $400 \times 200 \times 10$ grid points, and add a density grid of gaseous-dusty mixture with the dust properties as explained above. 

For all synthetic SEDs and images, the total flux density was determined via the integration over the whole source and then compared to observationally determined values in Table~\ref{tab_flux_density}.

\subsubsection{Different morphologies: source components}

At first, we performed radiative transfer calculations for different geometries of circumstellar envelopes to assess whether they can reproduce the detected high polarization degree in $K_{\rm s}$-band and the NIR-excess. First, we started with the simplest models with a star at the centre and flattened envelope and/or bow shock. For all these cases, the polarization degree remains below $10\%$. Breaking the spherical symmetry by introducing the cavities, having half-opening angle of $45^{\circ}$, increases the polarization degree to $\sim 10\%$ (without a bow shock layer). Adding a dense bow shock layer, the linear polarization fraction typically reaches $>20\%$, depending on the density of the envelope and the inclination. Qualitatively, the dependence of the linear polarization degree on the envelope geometry is sketched in Figure~\ref{fig_geometry_poldegree}.

The density grid in the radiative transfer models has different morphological and density components, whose characteristics are explained below:

\begin{itemize}
  \item {\bf star:} Based on the upper luminosity limit of $L_{\rm DSO} \lesssim 30\,L_{\odot}$ \citep{Eckart2013,Witzel2014}, the central star of the DSO should belong to the category of pre-main-sequence stars with the mass constraint $M_{\star} \lesssim 3 M_{\odot}$ \citep{2015wds..conf...27Z}. Given the NIR-excess, i.e. the rising SED longward of $2$ microns, it should belong to the category of class I objects -- protostars \citep{1987IAUS..115....1L}\footnote{The Lada-Wilking morphological classification of young stellar objects based on the spectral slope of their SEDs, $\alpha=\mathrm{d}\log{(\lambda F_{\lambda})}/\mathrm{d}\log{(\lambda)}$: class I sources ($0<\alpha\lesssim 3$), class II sources ($-2\lesssim \alpha \leq 0$), and class III sources ($-3<\alpha\lesssim -2$).} with an age of $10^5\,{\rm yr}$ up to $10^6\,{\rm yr}$. The comparison of stellar evolutionary tracks for different masses is in Fig.~\ref{fig_hr_dso} for the metallicity fraction of $Z=0.02$, which was constructed based on the computed tables by \citet{2000A&A...358..593S}. In the set of Monte Carlo simulations, we vary the luminosity and the temperature of stars to fit the observed flux density values and we get reasonable match for $M_{\star}=1.3\,M_{\odot}$, $L_{\star}=19.6\,L_{\odot}$ and $T_{\rm eff}=4220\,\rm{K}$ -- labelled as the red cross in Fig.~\ref{fig_hr_dso}. These stellar parameters were used in all the simulations presented in this paper, unless indicated otherwise.  
  
  \begin{figure}[tbh]
     \centering
     \includegraphics[width=0.5\textwidth]{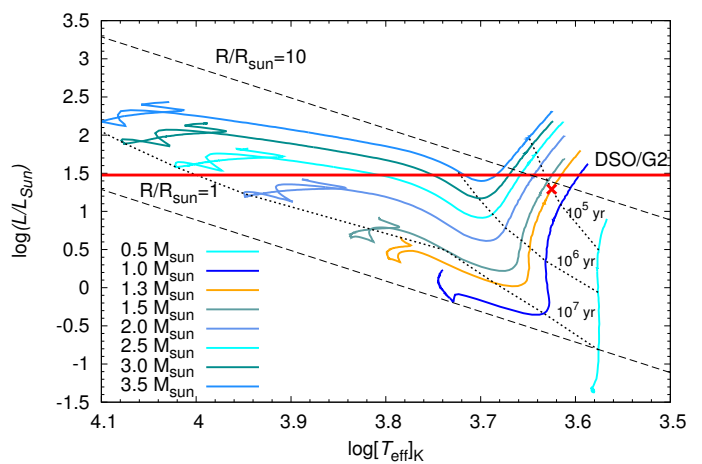}
     \caption{A set of the evolutionary tracks of pre-main-sequence stars based on \citet{2000A&A...358..593S}. The red line marks the upper limit for the bolometric luminosity of the DSO, $L_{\rm DSO} \lesssim 30\,L_{\odot}$. The orange solid line depicts the mass of $1.3\,M_{\odot}$, which we used for choosing the input parameters for radiative transfer calculations (red point).}
     \label{fig_hr_dso}
  \end{figure}
  
  \item {\bf flattened envelope:} To the first approximation, the dust-enshrouded star may be modelled as a star surrounded by a rotationally flattened, infalling dusty envelope that forms a disc at the corotation radius $r_{\rm c}$. The density profile is given by \citep{1976ApJ...210..377U},
\begin{align}
  \rho(r,\theta)=\frac{\dot{M}_{\rm acc}}{4\pi(G M_{\rm DSO}r_{\rm c}^3)^{1/2}}\left(\frac{r}{r_{\rm c}}\right)^{-3/2}\left(1+\frac{\mu}{\mu_0}\right)^{-1/2} \nonumber \\
  \times\left(\frac{\mu}{\mu_0}+\frac{2 \mu_0^2 r_{\rm c}}{r}\right)^{-1}\,,\label{eq_ulrich_density}
\end{align}  
   where $\dot{M}_{\rm acc}$ is the infall rate. The quantities $\mu$ and $\mu_0$ are related by an equation of the streamline, $\mu_0^3+\mu_0(r/r_{\rm c}-1)-\mu(r/r_{\rm c})=0$.
 
  \begin{figure*}[tbh]
     \centering
     \begin{tabular}{ccc}
     \includegraphics[width=0.33\textwidth]{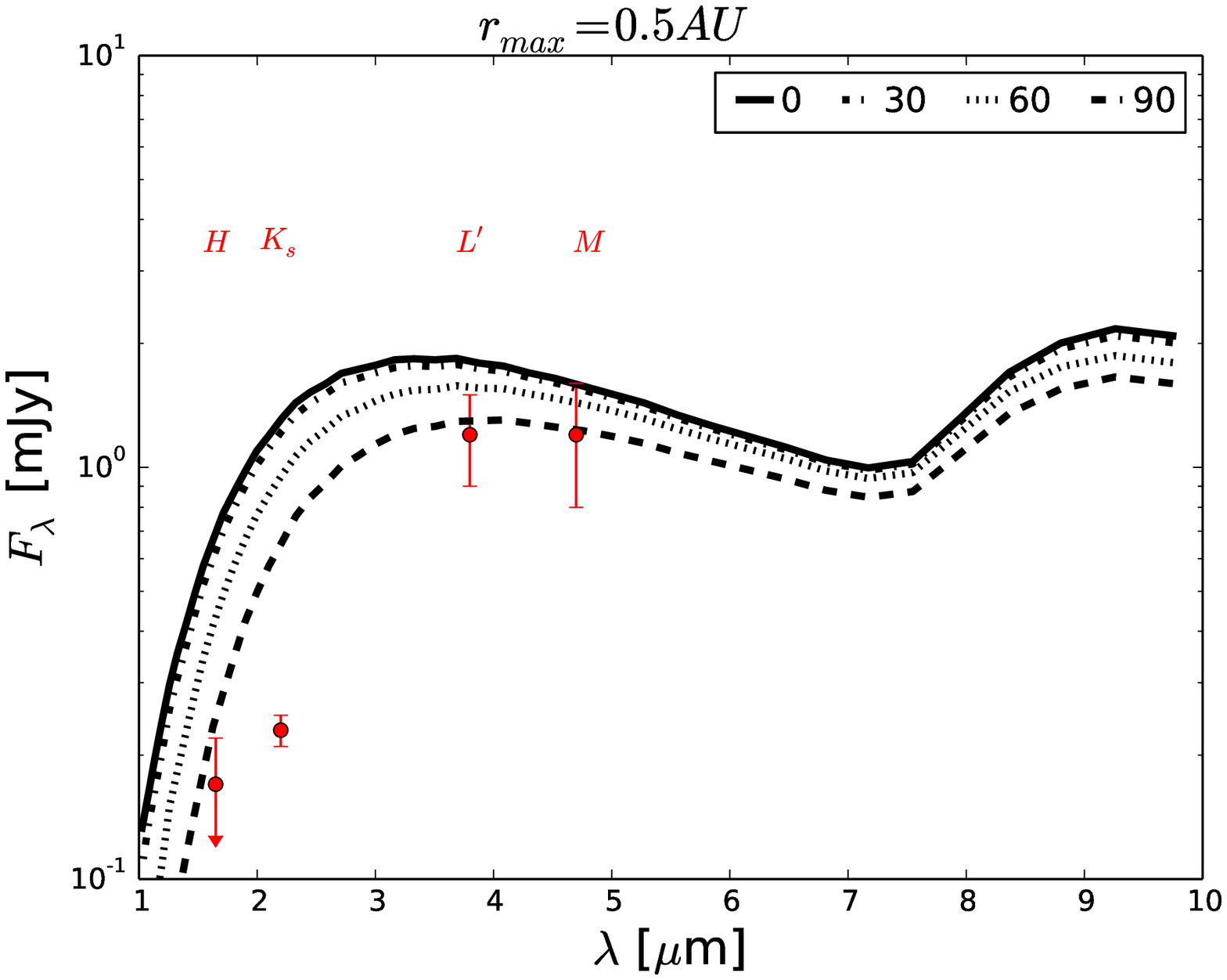} &\includegraphics[width=0.33\textwidth]{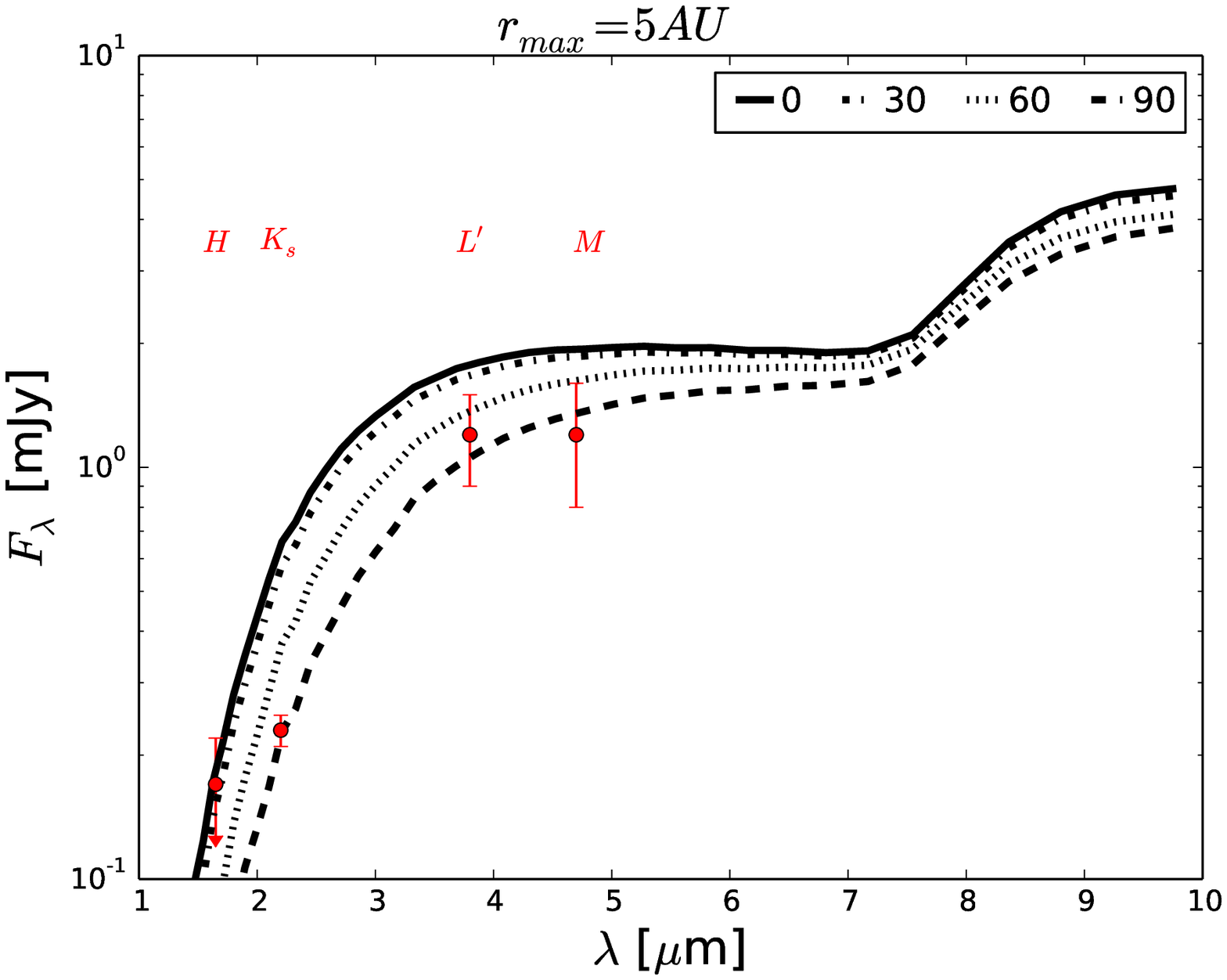} & \includegraphics[width=0.33\textwidth]{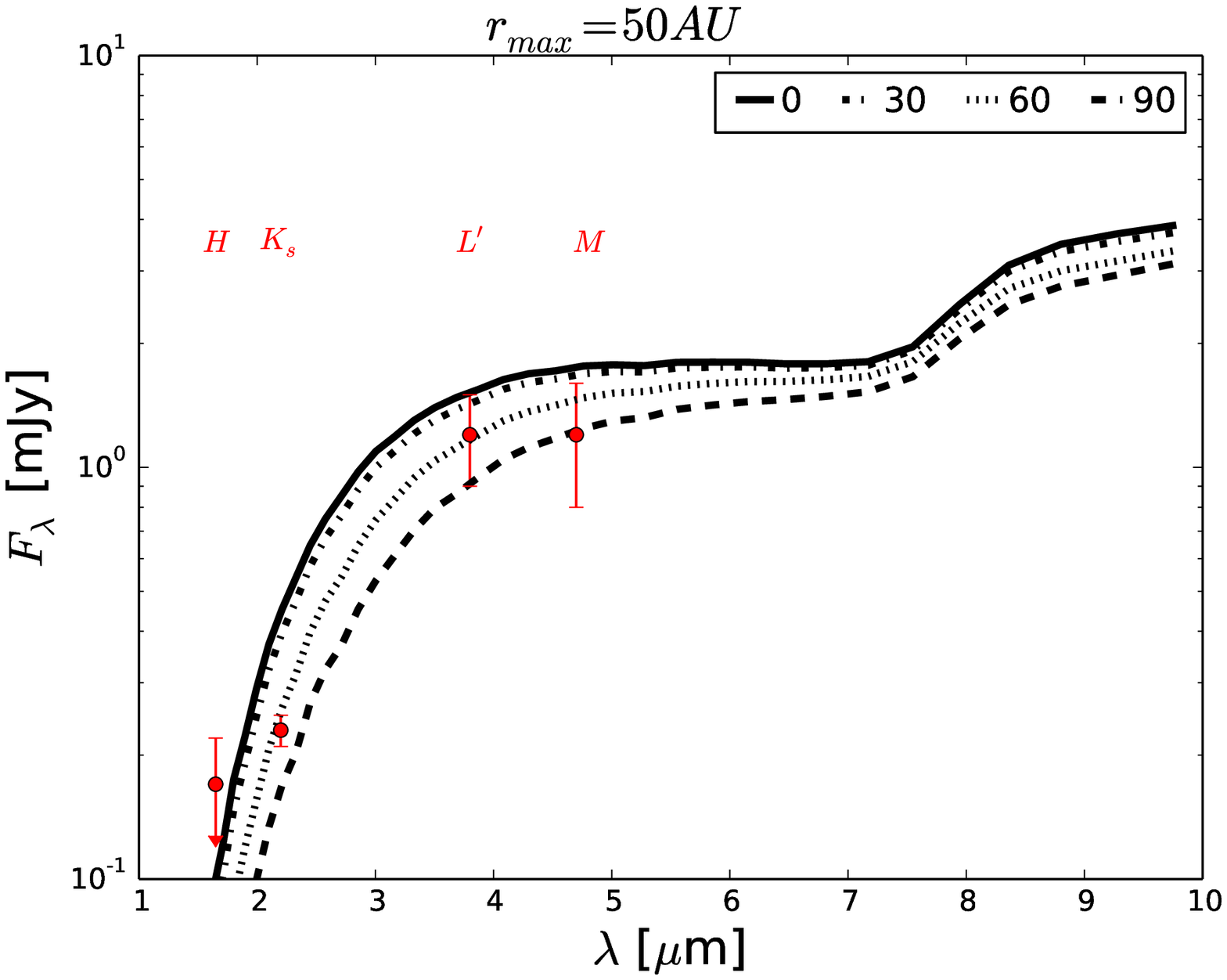}
     \end{tabular}
     \caption{Model SEDs of a star surrounded by rotationally flattened envelope for different inclinations in the range $(0^{\circ},90^{\circ})$ with an increment of $30^{\circ}$, see the key for different linestyles representing different inclinations. The points with error bars correspond to observationally inferred values, see Table~\ref{tab_flux_density}. \textbf{Left panel:} The maximum radius of the envelope is set to $r_{\rm max}=0.5\,{\rm AU}$. \textbf{Middle panel:} $r_{\rm max}=5\,{\rm AU}$. \textbf{Right panel:} $r_{\rm max}=50\,{\rm AU}$.}
     \label{fig_ulrich_sed}
  \end{figure*}
  
   \begin{figure*}[tbh]
     \centering
     \begin{tabular}{ccc}
     \includegraphics[width=0.33\textwidth]{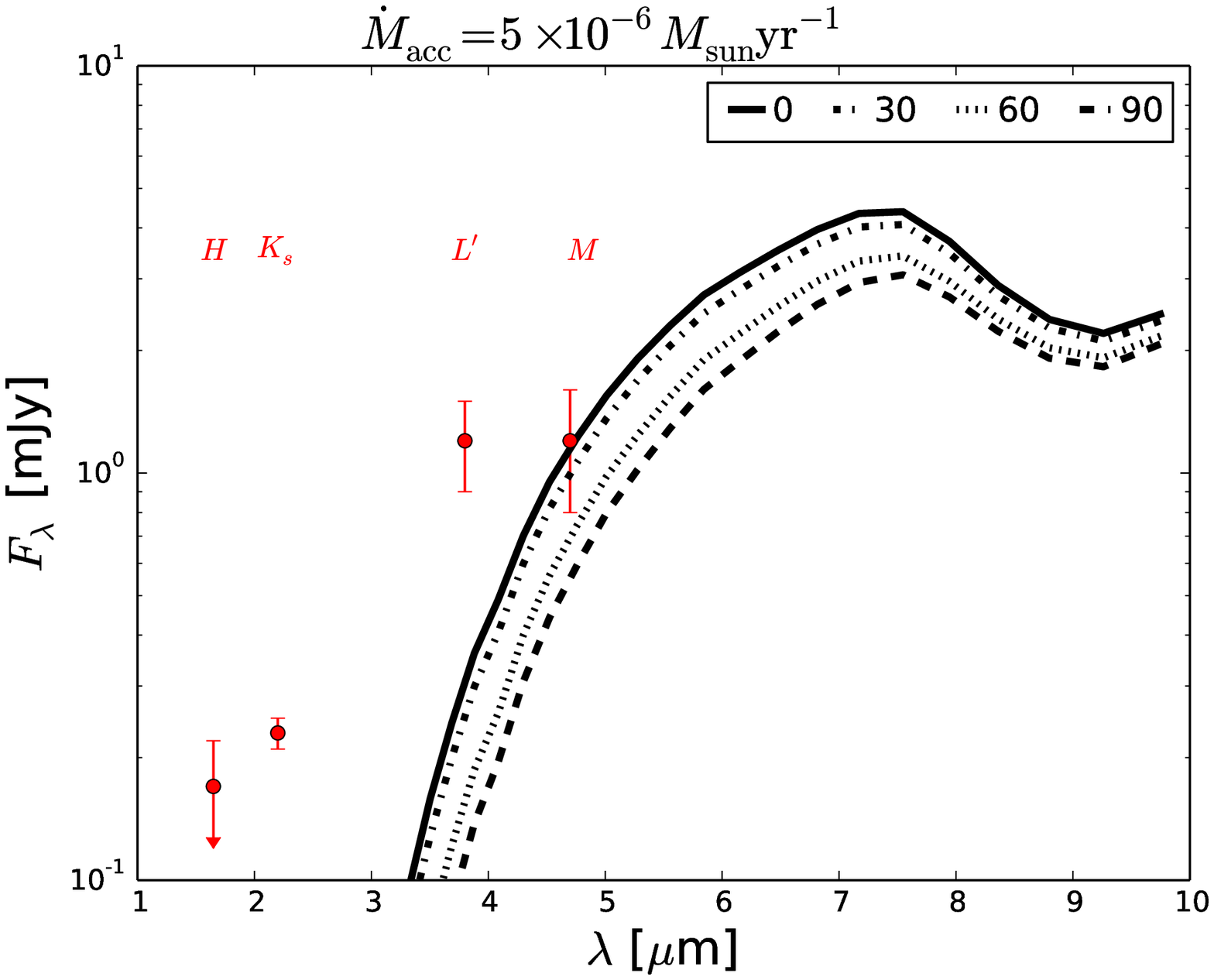} &\includegraphics[width=0.33\textwidth]{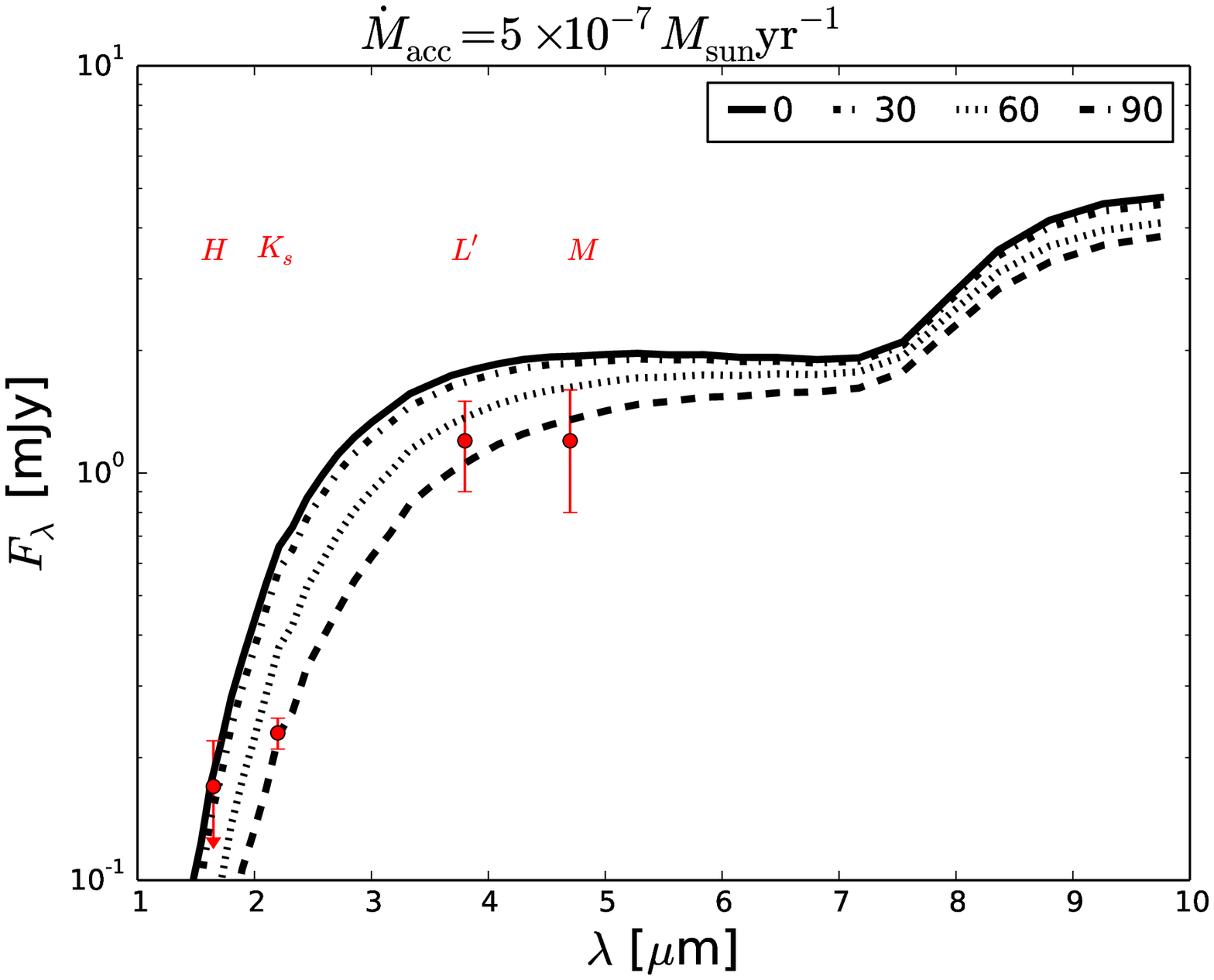} & \includegraphics[width=0.33\textwidth]{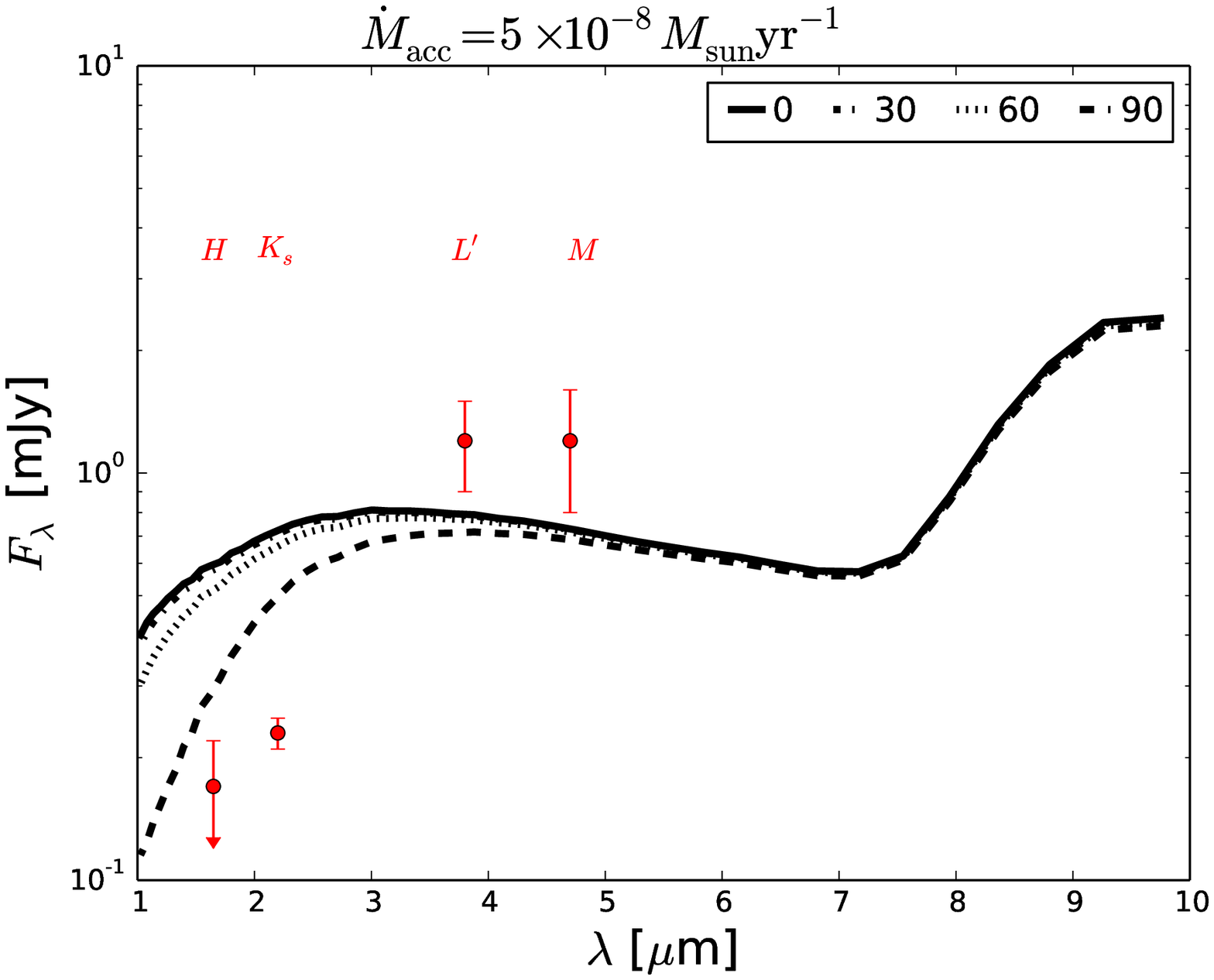}
     \end{tabular}
     \caption{Model SEDs of a star surrounded by rotationally flattened envelope with the fixed maximum radius of $r_{\rm max}=5\,\rm{AU}$ for different inclinations in the range $(0^{\circ},90^{\circ})$ with an increment of $30^{\circ}$, see the key for different linestyles representing different inclinations. The points with error bars correspond to observationally inferred values, see Table~\ref{tab_flux_density}. \textbf{Left panel:} The accretion rate is set to $\dot{M}_{\rm acc}=5\times 10^{-6}\,M_{\odot}{\rm yr^{-1}}$. \textbf{Middle panel:} $\dot{M}_{\rm acc}=5\times 10^{-7}\,M_{\odot}{\rm yr^{-1}}$. \textbf{Right panel:} $\dot{M}_{\rm acc}=5\times 10^{-8}\,M_{\odot}{\rm yr^{-1}}$.}
     \label{fig_ulrich_sed_accretion}
  \end{figure*}
  
  The model of Ulrich envelope can match the SED of the DSO for stellar parameters $M_{\star}=1.3\,M_{\odot}$, $L_{\star}=4.3\,L_{\odot}$, $T_{\rm eff}=4400\,{\rm K}$, $R_{\star}=3.3\,R_{\odot}$ (age $\sim 760\,000\,{\rm yr}$). The suitable parameters of the envelope are $\dot{M}_{\rm acc}=5\times 10^{-7}\,M_{\odot}\,{\rm yr^{-1}}$, $r_{\rm min}=10\,R_{\star}$, and $r_{\rm c}=20\,R_{\star}$, where $r_{\rm min}$ is the minimal distance of the envelope from the star. We vary the maximum distance of the envelope $r_{\rm max}$ from $0.5\,{\rm AU}$ up to $50\,{\rm AU}$, which affects the SED due to a different dust temperature distribution. For the SEDs of the flattened envelope at different inclinations $(0^{\circ}, 30^{\circ}, 60^{\circ}, 90^{\circ})$ and three different maximum radii $r_{\rm max}=[0.5, 5, 50]\,\rm{AU}$, see Fig.~\ref{fig_ulrich_sed}. Because of observational uncertainties, more suitable configurations are possible, e.g. $r_{\rm max}=5\,{\rm AU}$ and the inclination of $90^{\circ}$ or $r_{\rm max}=50\,{\rm AU}$ and the inclination in the range of $i=(60^{\circ},90^{\circ})$. The lower value of $r_{\rm max}=0.5\,{\rm AU}$ is not suitable because of the large flux in $K_{\rm s}$ band, which can be explained by a lower extinction and a more prominent stellar emission. In the range $r_{\rm max}=(5,50)\,\rm {AU}$, the SED does not depend much on this parameter because of the decreasing gas and dust density according to Eq.~\eqref{eq_ulrich_density}. 
  On the other hand, the SED depends strongly on the accretion rate $\dot{M}_{\rm acc}$. We vary the accretion rate between $\dot{M}_{\rm acc}=(5\times 10^{-6}, 5\times 10^{-8})\,M_{\odot}\,{\rm yr^{-1}}$ for the fixed maximum radius of $r_{\rm max}=5\,{\rm AU}$, see Fig.~\ref{fig_ulrich_sed_accretion}. Clearly, the best match of SED values is for an intermediate value of $\dot{M}_{\rm acc}=5\times 10^{-7}\,M_{\odot}\,{\rm yr^{-1}}$ (Middle panel), increasing or decreasing the accretion rate by an order of magnitude significantly changes calculated fluxes, which is caused by large changes in the dust density, $\rho \propto \dot{M}_{\rm acc}$, see Eq.~\eqref{eq_ulrich_density}.   
   Although a star surrounded by a flattened envelope can satisfactorily explain the SED of the DSO, the calculated polarized emission in $K_{\rm s}$ band from radiative transfer models yields the maximum total polarization degree of $\overline{p}_{\rm K,L}=2.7\%$ for $i=90^{\circ}$, i.e. well below the detected value of $\sim 30\%$ \citep{Shahzaman2016}. This implies that the simple geometry of the flattened envelope cannot alone explain the properties of the DSO.

  \item {\bf spherically concentrated dusty envelope/flared disc:} For more complex models, we set up a stratified spherical dusty envelope that was intersected by bipolar cavities in some models with smaller number densities (see below). In the final set of models, the spectral slope could be reproduced by the following mean mass densities of the gas$+$dust mixture: $1 \times 10^{-13}\,{\rm g\,cm^{-3}}$ for $r=0.1\,{\rm AU}$--$1.0\,{\rm AU}$, $1 \times 10^{-14}\,{\rm g\,cm^{-3}}$ for $r=1.0\,{\rm AU}$--$2.0\,{\rm AU}$, and $1 \times 10^{-15}\,{\rm g\,cm^{-3}}$ for $r=2.0\,{\rm AU}$--$3.0\,{\rm AU}$.  
  \item {\bf bipolar cavity:} The bipolar cavities were introduced in the model to increase the non-spherical nature of the DSO, which is naturally needed to obtain a non-zero total polarization degree in the model, see Fig.~\ref{fig_geometry_poldegree}. Furthermore, the walls of cavities also provide effective space for single scattering events of stellar and dust photons that escape from the star and the envelope, respectively. The density inside cavities is set to be uniform and smaller than in the envelope by several orders of magnitude: $1 \times 10^{-20}\,{\rm g\,cm^{-3}}$, which is a typical order of magnitude \citep[see e.g.][]{Sanchez2016}. The half-opening angle of the cavity is initially varied to investigate the impact of the parameter space on the SED and the polarization, then set to $\theta_0=45^{\circ}$.   
  
  \begin{figure*}[tbh]
  \centering
  \begin{tabular}{cc}
  \includegraphics[width=0.5\textwidth]{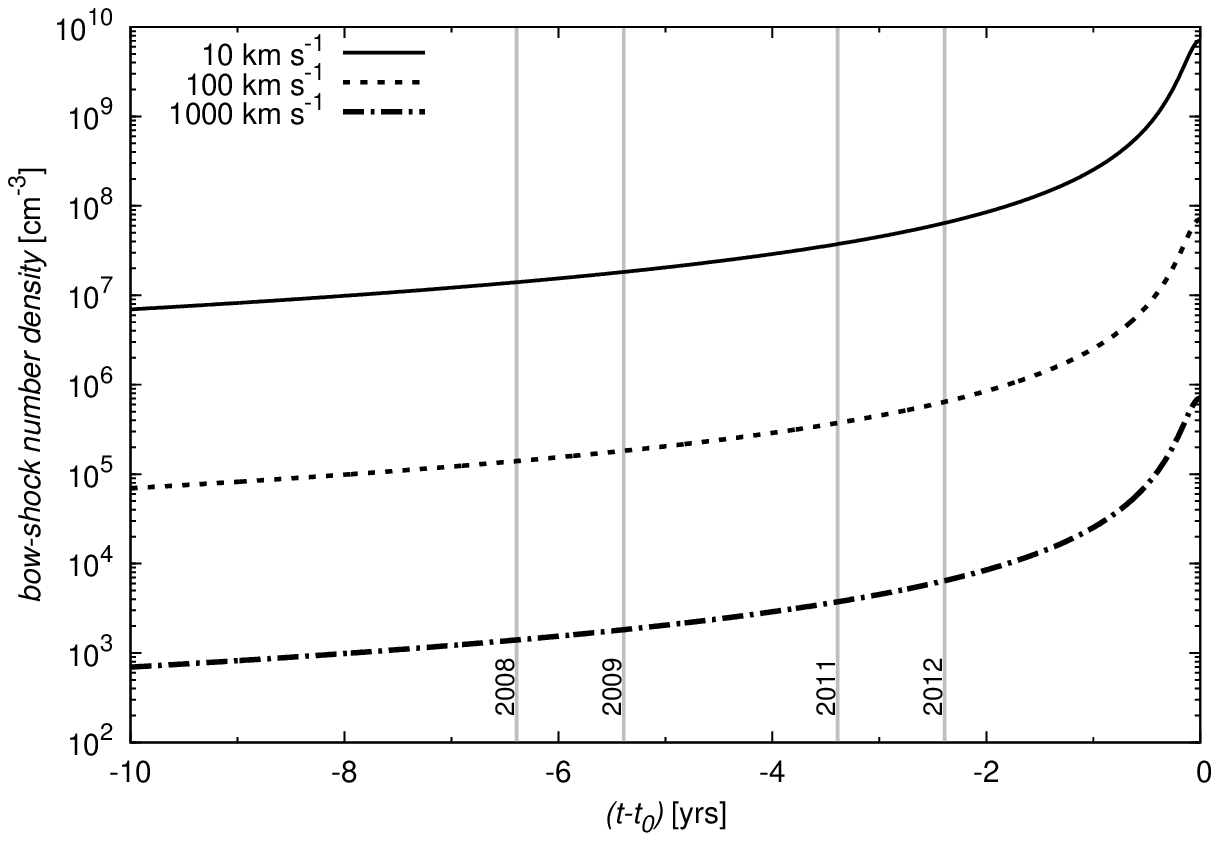} & \includegraphics[width=0.5\textwidth]{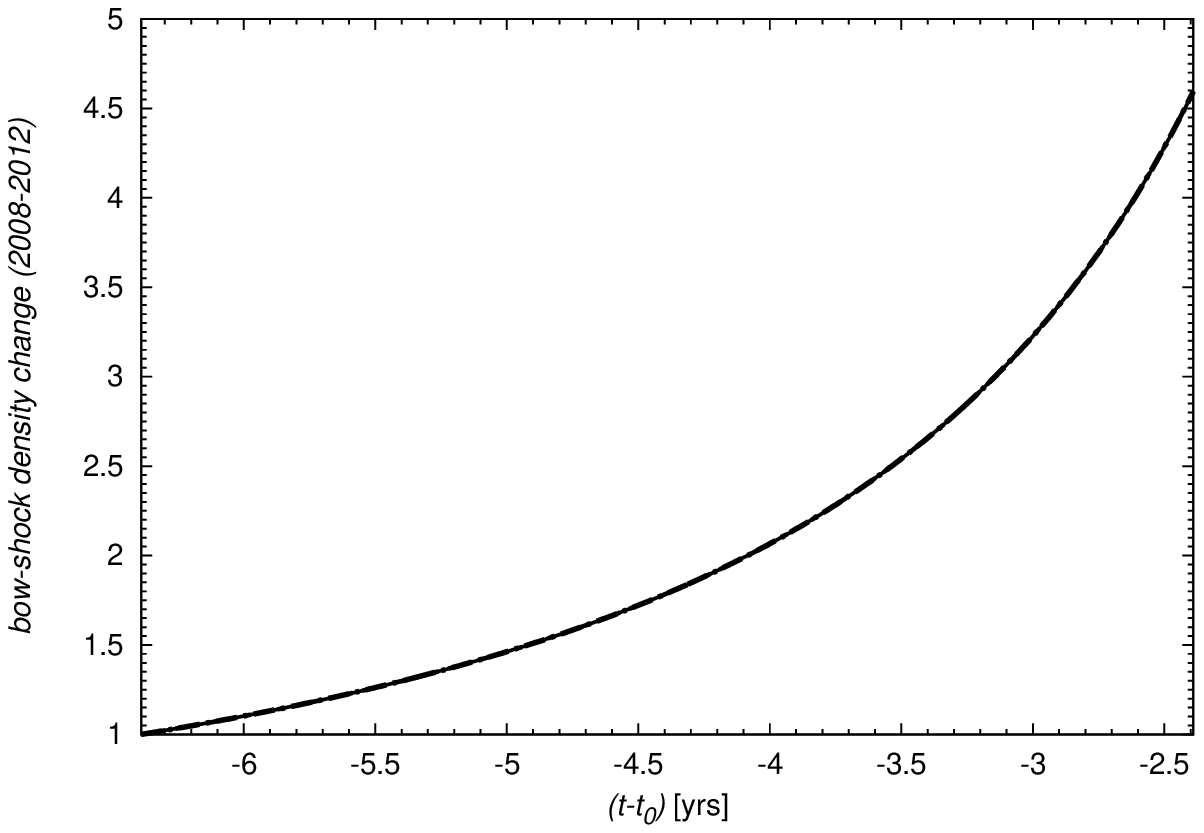}
  \end{tabular}
  \caption{\textbf{Left:} The number density of the shocked layer for different wind velocities (see the key) and the stellar mass-loss rate of $\dot{m}_{\rm w}=10^{-8}\,{\rm M_{\odot}\,yr^{-1}}$. \textbf{Right:} The relative change in the bow-shock density between the epochs $2008$--$2012$.}
  \label{fig_bowshock_density}
\end{figure*} 
  
  \item {\bf bow shock:} The DSO and other objects in the S-cluster are expected to move supersonically through the ambient medium, especially close to the pericentre of their orbits \citep[see Fig.~1 in][]{zajacek2016}, see also \citet{de-colle2014,Ballone2013,Ballone2016,2016MNRAS.459.2420C}. The expected Mach number of the DSO is $M=v_{\rm rel}/c_{\rm s} \lesssim 10$ \citep{zajacek2016}, where $v_{\rm rel}$ is the relative velocity of the DSO with respect to the medium and $c_{\rm s}$ is the local sound speed. This necessarily leads to the formation of the bow shock layer whose densest part lies ahead of the star close to the stagnation point. It was shown that the bow shock of the DSO can contribute to the continuum as well as line emission of the source \citep{Scoville2013}. The contribution of the bow-shock NIR continuum emission depends on the dust content and dust size distribution in the bow shock layer, which by itself is not a trivial hydrodynamical problem \citep{2011ApJ...734L..26V}. However, the existence of a bow shock would further increase the overall asymmetry of the source and hence make the total linear polarization degree larger, see Fig.~\ref{fig_geometry_poldegree}. 
  
  For modelling the stellar bow shock associated with the supersonic motion of the DSO, we apply the analytical model of \citet{wilkin1996} for the shape of the axisymmetric layer,
  \begin{equation}
     R(\theta)=R_0 \csc{\theta} \sqrt{3(1-\theta \cot{\theta})}\,,
     \label{eq_bowshock_shape}
 \end{equation}   
 where $\theta$ is a polar angle from the axis of symmetry as seen from the star that lies at the coordinate origin. The standoff distance $R_0$ depends on two intrinsic stellar parameters -- the mass-loss rate $\dot{m}_{\rm w}$ and the terminal wind velocity $v_{\rm w}$ -- as well as the density profile of the ambient medium $\rho_{\rm{a}}(r)$ and the relative velocity of the star with respect to the ambient medium $v_{\rm rel}$. The general form for the standoff distance is \citep{wilkin1996,1997ApJ...474..719Z,2016MNRAS.459.2420C},
  \begin{equation}
     R_0=\left(\frac{\dot{m}_{\rm w} v_{\rm w}}{\Omega (1+\beta) \rho_{\rm a} v_{\rm rel}^2}\right)^{1/2}\,,
     \label{eq_standoff}
  \end{equation}
where $\beta$ is the ratio of the thermal and the ram pressure of the ambient medium, $\beta=P_{\rm th}/P_{\rm ram}$. When a star moves supersonically, its Mach number is larger than one, $M=v_{\rm rel}/c_{\rm s}=1/\sqrt{\kappa \beta}>1$, where $\kappa$ is an adiabatic index. When the thermal pressure is low, i.e. the ram pressure is much higher than the thermal pressure of the medium, the term $\beta \rightarrow 0$. For the ambient medium, we assume the radial profile as in  \citet{zajacek2016},  
\begin{align}
n_{{\rm a}} &\approx n_{{\rm a}}^0\left(\frac{r}{r_{0}}\right)^{-\gamma_n}\,, \label{eq_density}\\
T_{{\rm a}} &\approx T_{{\rm a}}^0\left(\frac{r}{r_{0}}\right)^{-\gamma_T}\,, \label{eq_temperature}  
\end{align}
where we set $r_0$ to the Schwarzschild radius $r_{\rm s}$ ($r_{{\rm s}} \equiv 2GM_{\bullet}/c^2 \doteq 2.95\times 10^5\,M_{\bullet}/M_{\odot}\,{\rm cm}$). The normalisation parameters are $n_{{\rm a}}^{0}=1.3\times 10^7\,{\rm cm}^{-3}$ and $T_{{\rm a}}^0=9.5 \times 10^{10}\,{\rm{K}}$ and the power-law indices are $\gamma_{n} \approx \gamma_{T}=1$. At the pericentre, the estimated distance of the DSO according to \citet{Valencia2015} is $r_{\rm P}=a(1-e)\approx 1925\,r_{\rm s}$ and the corresponding orbital velocity is $v_{\rm P}\approx 6600\,{\rm km\,s^{-1}}$. If we assume $v_{\rm rel}\approx v_{\rm P}$ and $\kappa \approx 1$, then the factor $\beta\approx 0.01$, so the thermal pressure may be neglected at the pericentre \footnote{However, at the apocentre of the DSO orbit, the factor $\beta \approx 0.9$, since the ratio of the thermal factors at the apocentre and the pericentre may be expressed as $\beta_{\rm A}/\beta_{\rm P}=(1+e)/(1-e)$.}.

 The factor $\Omega$ is the solid angle, into which the stellar wind/outflow is blown. For a general case, the factor $\Omega=2\pi(1-\cos{\theta_0})$, where $\theta_0$ is a half-opening angle of the outflow. For an isotropic stellar wind, we have $\theta_0=\pi$, and so we naturally get $\Omega=4\pi$. For bipolar cavities with $\theta_0=\pi/4$, we get $\Omega=2\pi(1-\sqrt{2}/2)\approx 1.84$.
 
 The density of the shocked wind layer $\rho_{\rm bs}$ may be estimated from the Rankine-Hugoniot jump conditions \citep{2016MNRAS.459.2420C},

 \begin{equation}
  \rho_{\rm bs}=\rho_{\rm a} \frac{\kappa+1}{\kappa-1}\left(\frac{v_{\rm rel}}{v_{\rm w}}\right)^2 (1+\beta)\,.
  \label{eq_bowshock_density}
 \end{equation}
 In case the motion of the ambient medium may be neglected, $v_{\rm rel}=v_{\rm \star}$, and the thermal pressure is negligible (close to the pericentre), $\beta \rightarrow 0$, then the relative change in density of the shocked layer between the years $2008$-$2012$ (or any two points at the distance of $r_0$ and $r'$ along the orbit) may be expressed as,
 \begin{equation}
   \frac{\rho'}{\rho_0}=\left(\frac{r_0}{r'}\right)^{(\gamma_{n}+1)}\left(\frac{2a-r'}{2a-r_0}\right)\,,
   \label{eq_relchange}
 \end{equation}
 which for $\gamma_{n}=1$ and the orbital elements according to \citet{Valencia2015} yields $\rho_{2012}/\rho_{2008}\approx 4.35$. The number density of the shocked layer as function of time as well as the relative change between the years (2008-2012) are plotted in Fig.~\ref{fig_bowshock_density} in left and right panels, respectively. For these calculations, the thermal term was included according to Eq.~\eqref{eq_bowshock_density}, which explains a small difference with respect to the estimate of the relative density increase above. The stellar parameters were adopted from the previous analysis of \citet{Scoville2013} and \citet{zajacek2016}: the mass-loss rate $\dot{m}_{\rm w}=10^{-8}\,{\rm M_{\odot}\,yr^{-1}}$ and variable wind velocities in the range $10\,{\rm km\,s^{-1}}$--$1000\,{\rm km\,s^{-1}}$. For the radiative transfer simulations, we tried different values of the bow-shock density and the best match to the observed SED and the polarized emission was reached for $v_{\rm w}=10\,{\rm km\,s^{-1}}$, i.e. slow outflow.
 \end{itemize} 

 The summary of the polarization degree and the source colour $(K_{\rm s} - L')$ (with and without line-of-sight extinction) for all the circumstellar geometries, which were tested, is summarized in Table~\ref{tab_geometries}.

\begin{table*}
{\Large
   \resizebox{\textwidth}{1.5cm}{  
  \begin{tabular}{c|c|c|c}
  \hline
  \hline
  \textbf{Geometry} & $K_{\rm s}$-band Total Linear Polarization Degree $\overline{p}_{\rm K,L}$ [\%] & $(K_{\rm s}-L')_{\rm int}$ (intrinsic) & $(K_{\rm s}-L')_{\rm ext}=(K_{\rm s}-L')_{\rm int}+(A_{K}-A_{L})$ (with extinction)\\
  \hline
  Star  & $0$ ($\sim 6\%$ foreground pol. at Gal. Ctr.) & $-0.9$ & $0.4$  \\
  Star+rotationally flattened envelope ($90^{\circ}$ inclination) & $2.7$ & $1.6$ & $2.9$  \\
  Star+flared disc & $3.2$ & $0.3$ & $1.6$\\
  Star+dense bow shock (inclined) & $4.1$ & $1.9$ & $3.2$\\
  Star+spherical dusty envelope+dense bow shock (inclined) & $1.0$ & $1.6$ & $2.9$\\
  Star+flattened envelope+cavities ($90\%$ inclination) & $10.1$ & $1.2$ & $2.4$\\
  Star+flared disc+flattened envelope+cavities ($90\%$ inclination) & $10.5$ & $-1.3$ & $-0.03$\\
  \rowcolor{Gray}
  \textbf{Star+envelope+cavities+dense bow shock} & $\mathbf{23.0}$ & $1.6$ & $\mathbf{2.9}$\\
  \hline
  \rowcolor{LightCyan}
  \textbf{Observations} \citep{Shahzaman2016} & $\mathbf{\sim 30}$  & $1.8$ & $\mathbf{3.1}$ \\
  \hline 
  \end{tabular}
  }}
  \caption{Circumstellar geometries with a different set of components. Important parameters are the total linear polarization degree in $K_{\rm s}$ band $(2.2\,\rm{\mu m})$ and the colour index $K_{\rm s}-L'$ (with and without line-of-sight extinction). The observed values for the polarization degree and the colour index are also included. The observed values of the polarization degree and the colour index are matched best by the composite model star+disc+cavities+dense bow shock.}
  \label{tab_geometries}
\end{table*}

\subsubsection{Final model: supersonic, dust-enshrouded star with non-spherical envelope}   

\begin{figure}[tbh]
  \centering
  \includegraphics[width=0.5\textwidth]{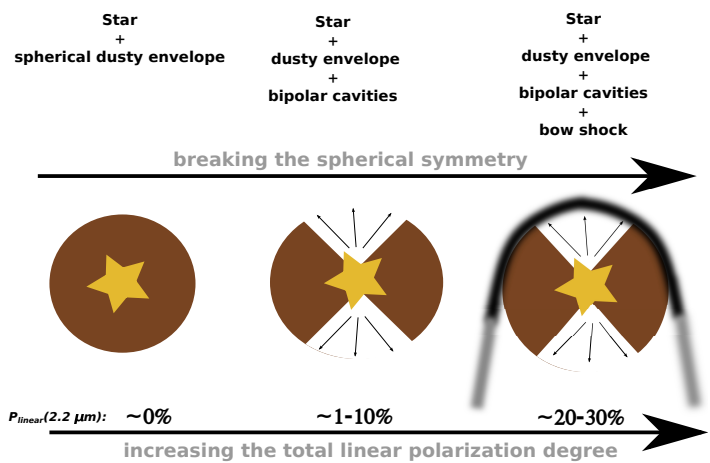}
  \caption{Sketch of how the geometry of circumstellar envelope affects the total linear polarization degree in $K_{\rm s}$ band.}
  \label{fig_geometry_poldegree}
\end{figure}

\begin{figure*}[tbh]
  \centering
  \begin{tabular}{cc}
  \includegraphics[width=0.5\textwidth]{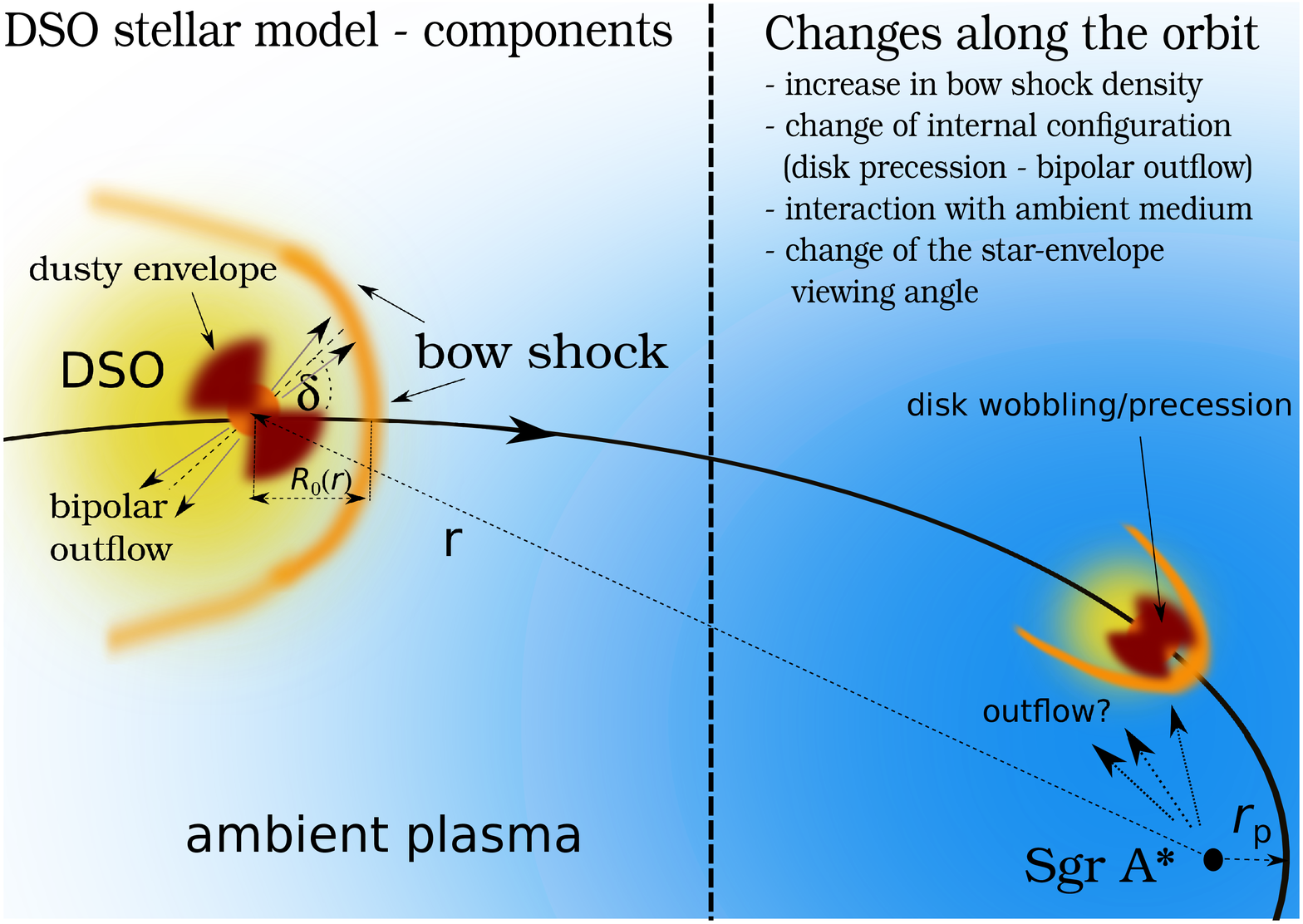} & \includegraphics[width=0.5\textwidth]{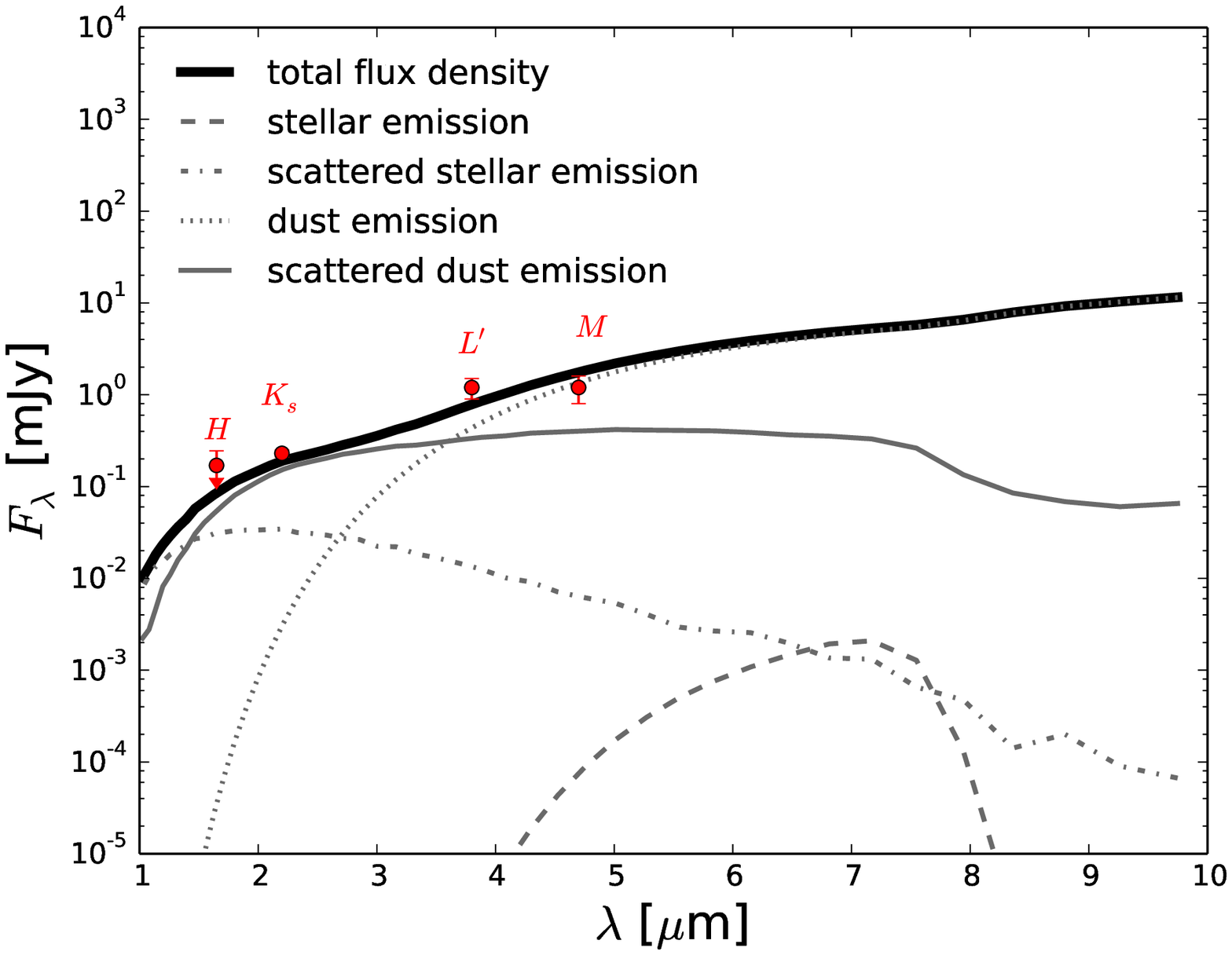}
  \end{tabular}
  \caption{\textbf{Left:} Illustration of the components of the DSO model explained as a pre-main-sequence star. The right side explains possible sources of the changes in the continuum emission of the DSO/G2 (polarization degree and angle). \textbf{Right:} The calculated SED for the composite model of the DSO (star, dusty envelope, bipolar cavities, bow shock) for the viewing angle of $90^{\circ}$. The thick solid line stands for the total continuum flux density, whereas gray lines represent individual source contributions (see the key). The points represent observationally inferred flux densities/limits for $H$, $K_{\rm s}$, $L'$, and $M$ bands.}
  \label{fig_dso_model}
\end{figure*}

From the set of radiative transfer simulations with different circumstellar geometries (see Table~\ref{tab_geometries}), the model that can meet all constraints listed in Table~\ref{tab_constraints} consists of the following components:
\begin{itemize}
  \item a pre-main-sequence star,
  \item spherically concentrated dusty envelope/geometrically thick disc,
  \item bipolar cavity with a half-opening angle $\theta_0=45^{\circ}$,
  \item bow shock,
\end{itemize} 
The detailed discussion of the adopted parameters and densities of the gas/dust envelope is in the previous subsection. The illustration of the model is in Fig.~\ref{fig_dso_model} (left panel). An important parameter in terms of the intrinsic geometry is the angle $\delta$ that determines the orientation of the bipolar outflow with respect to the axis of symmetry of the bow shock. Unless otherwise indicated, we set $\delta=0^{\circ}$, i.e. the bipolar cavities are aligned with the symmetry axis of the bow shock.

\begin{figure*}[tbh]
  \centering
  \begin{tabular}{cc}
  \includegraphics[width=0.5\textwidth]{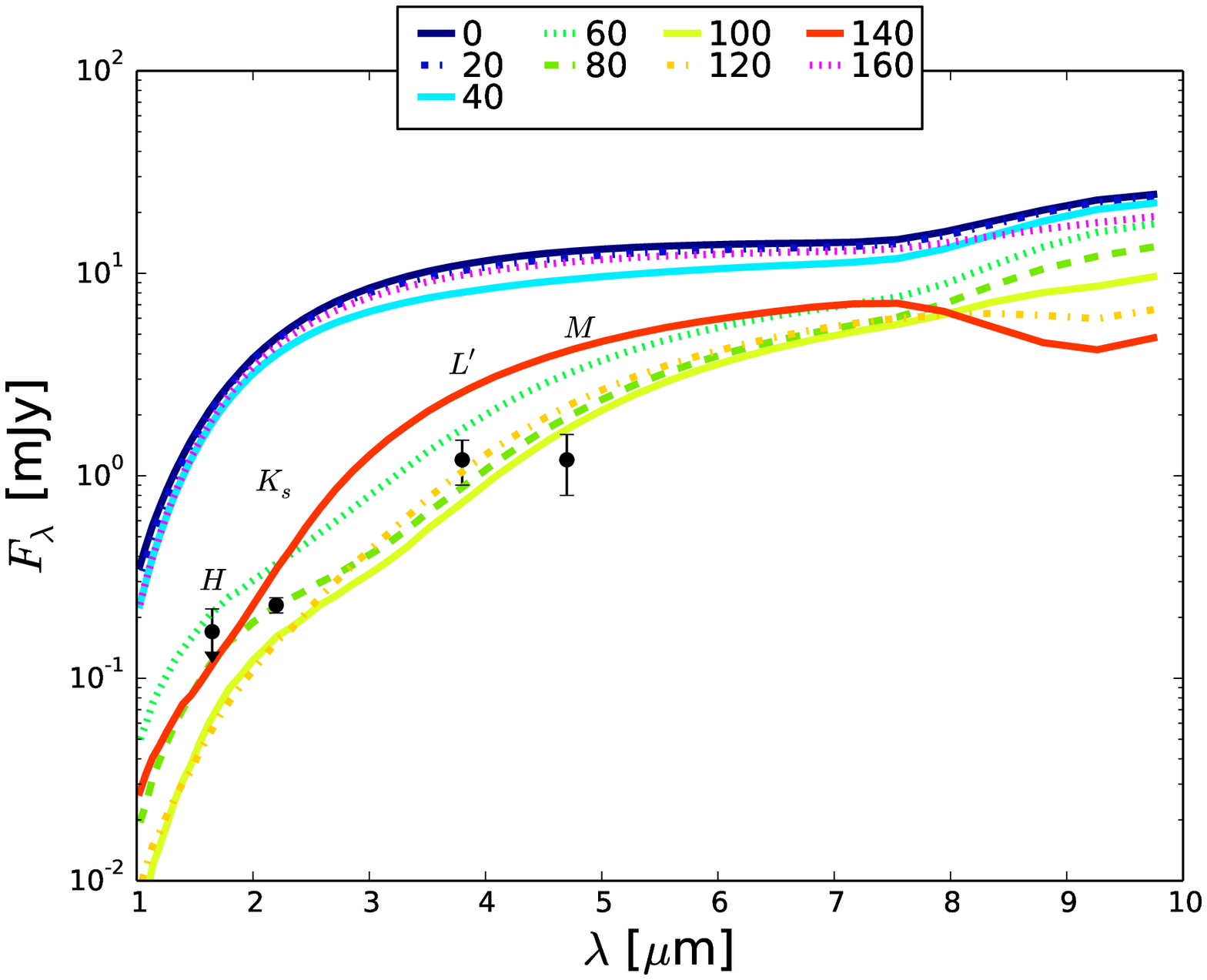} & \includegraphics[width=0.5\textwidth]{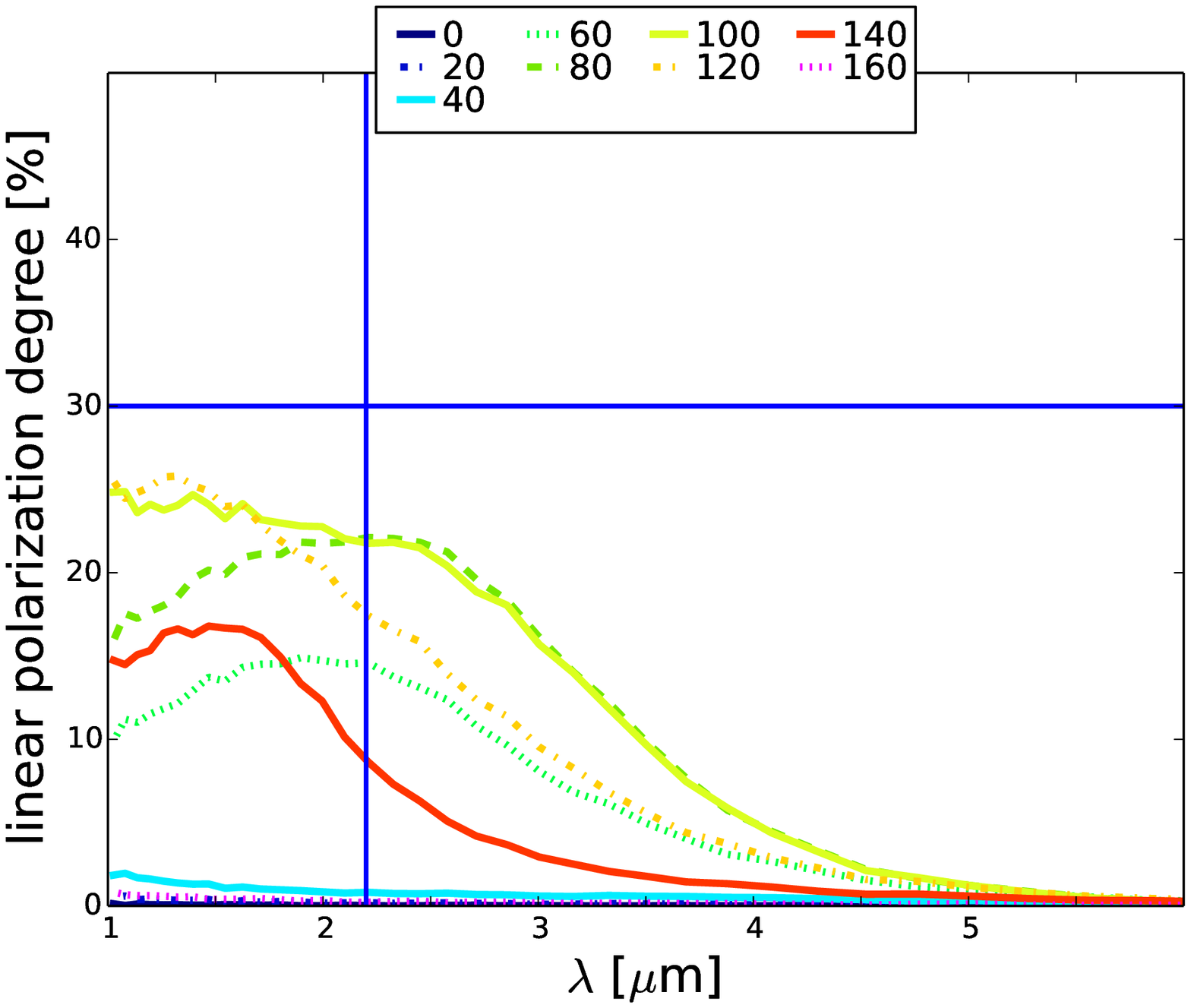}
  \end{tabular}
  \caption{\textbf{Left:} The spectral energy distribution of the composite stellar model of the DSO as function of the viewing angle. The points represent observationally inferred values. Different viewing angles from $0^{\circ}$ up to $180^{\circ}$ are labelled by different colours according to the key. \textbf{Right:} The linear polarization degree as function of wavelength for different viewing angles (see the key). The vertical thick line marks the values along the $K_{\rm s}$ band $(2.2\,{\mu m})$, the horizontal line denotes the value of $p_{\rm L}=30\%$, which is close to observationally inferred values for four consecutive epochs (2008, 2009, 2011, 2012; \citeauthor{Shahzaman2016}, \citeyear{Shahzaman2016}).}
  \label{fig_dso_sed_inc}
\end{figure*}

An exemplary model SED is in Fig.~\ref{fig_dso_model} (right panel) calculated for the viewing angle of $90^{\circ}$. Here we use the same convention for the viewing angle as \citeauthor{muzic2010}, \citeyear{muzic2010}, see their Fig. 3 -- $0^{\circ}$ corresponds to the front view of the bow shock (circular shape in projection), $90^{\circ}$ corresponds to the side view (bow-shock shape), and $180^{\circ}$ corresponds to the view from the tail part (also circular shape in projection). The thick black solid line represents the total continuum flux density and thinner, gray lines stand for individual components: stellar emission, scattered stellar emission, dust emission, and scattered dust emission. It is clearly visible that $L'$-band flux density is dominated by the thermal dust emission (direct photons), whereas $K_{\rm s}$-band emission is dominated by scattered emission, mainly scattered dust photons and to a smaller extent, scattered stellar photons. The stellar photospheric emission is negligible across the whole NIR- and MIR-spectrum. 


\begin{figure*}[tbh]
  \centering
  \begin{tabular}{cc}
  \includegraphics[width=0.5\textwidth]{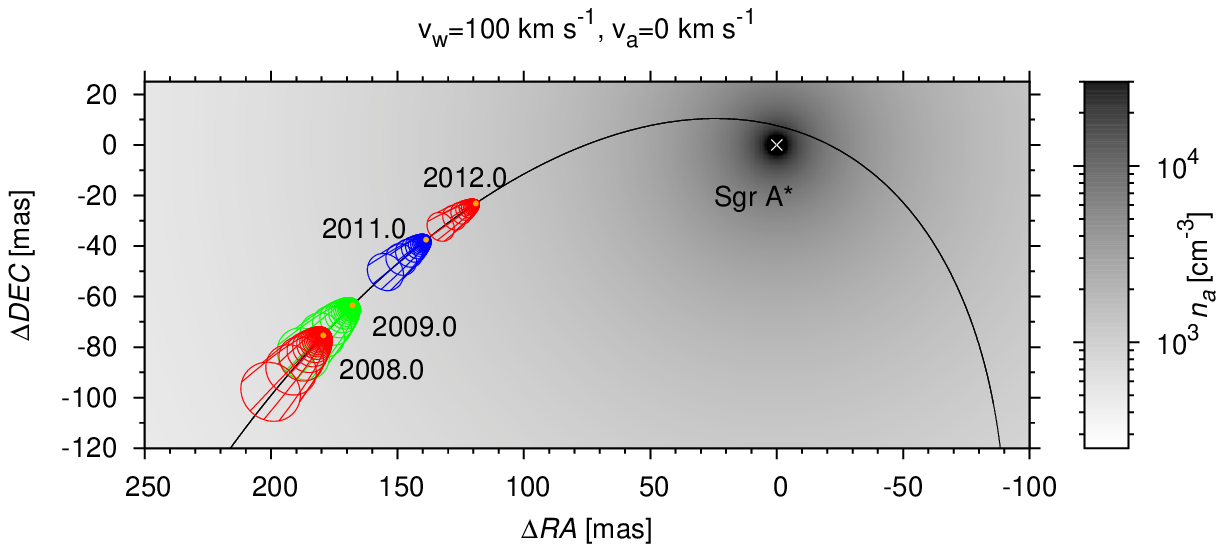} & \includegraphics[width=0.5\textwidth]{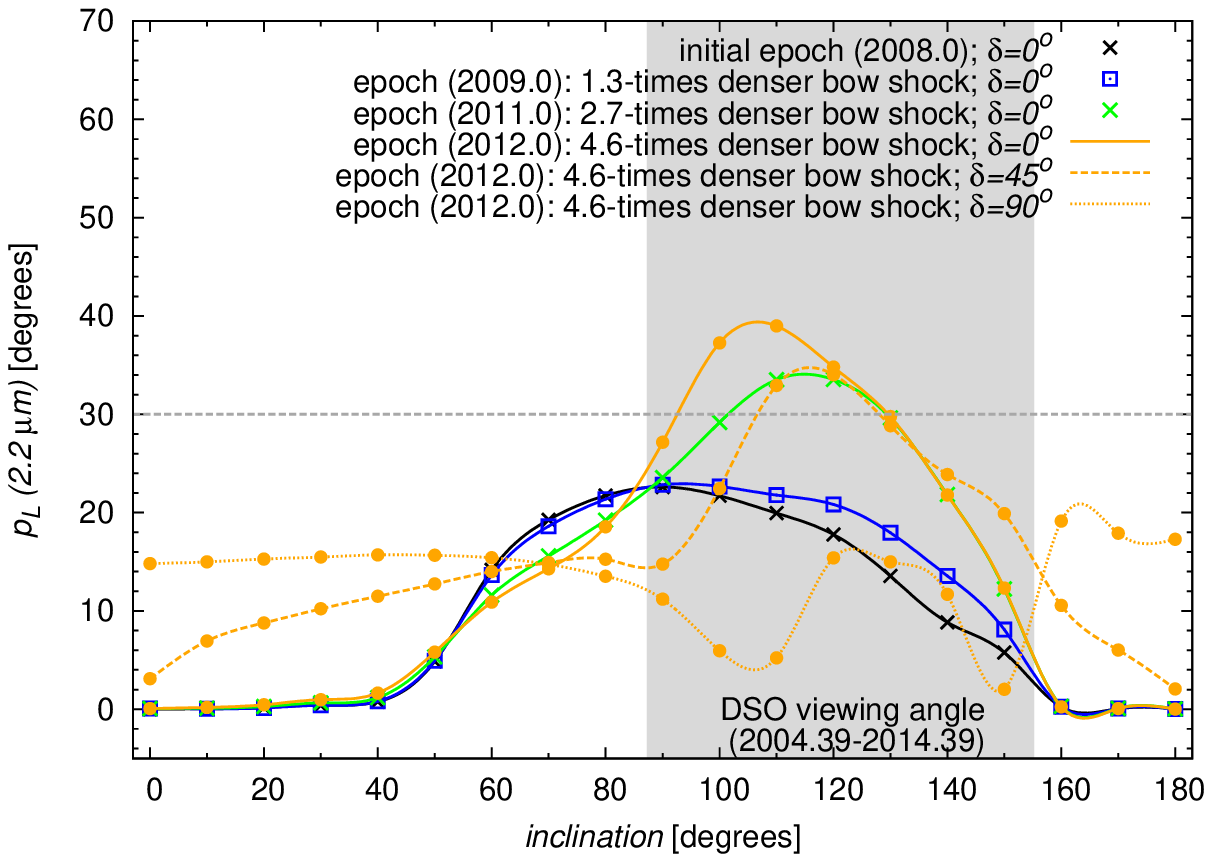}
  \end{tabular}
  \caption{\textbf{Left:} Schematic plot of the bow-shock evolution of the DSO along the orbit. The mass-loss rate was taken to be $\dot{m}_{\rm w}=10^{-8}\,{\rm M_{\odot}\,yr^{-1}}$ and the terminal wind velocity is $v_{\rm w}=100\,{\rm km\,s^{-1}}$. The ambient density was colour-coded according the distribution expressed by Eq.~\ref{eq_density}. \textbf{Right:} Total linear polarization degree in $K_{\rm s}$ band for different inclinations ($0$--$180$ degrees) of the DSO composite model. According to the key, the solid lines represent different epochs, 2008--2012, with a gradually increasing bow-shock density (see also Fig.~\ref{fig_bowshock_density}). The three orange lines associated with the largest bow-shock density represent the set-ups for three different position angles of the bipolar outflow, $\delta=0^{\circ}$, $45^{\circ}$, and $90^{\circ}$. The shaded rectangular region represents different angles of the bow-shock axis with respect to the line of sight for the period of 10 years before the pericentre passage of the DSO, assuming a negligible motion of the ambient medium in comparison with the orbital velocity of the DSO. The horizontal dashed line marks the polarization degree value of $p_{\rm L}=30\%$, which is approximately the observationally inferred degree for the DSO \citep{Shahzaman2016}.}
  \label{fig_linpol_inc}
\end{figure*}

We also check the dependence of the SED on the viewing angle, see Fig.~\ref{fig_dso_sed_inc} (left panel). Quantitatively, the best match is for viewing angles larger than $50^{\circ}$ and less than $130^{\circ}$. The dependence of the SED on the viewing angle implies a possible source of continuum variability as the DSO source orbits Sgr~A*. For $K_{\rm s}$ and $L'$ bands, the variability is a few $0.1\,{\rm mJy}$ within an expected viewing angle $\sim 80^{\circ}$--$150^{\circ}$, see Appendix~\ref{appa1} for estimates, which depend mainly on the relative velocity of the star with respect to the medium, which is in general uncertain. However, close to the pericentre, the relative velocity should approach the orbital velocity. 

The linear polarization degree as function of wavelength is depicted in Fig.~\ref{fig_dso_sed_inc} (right panel). The largest total polarization degree, $P_{\rm L}\gtrsim 20\%$, is for the viewing angles close to $90^{\circ}$ when the source is highly non-spherical. On the other hand, for the viewing angle either close to $0^{\circ}$ or $180^{\circ}$, the total linear polarization degree is $\lesssim 10\%$, since the source appears rather circular.

\begin{figure*}[tbh]
 \centering
 \begin{tabular}{c}
   \includegraphics[width=\textwidth]{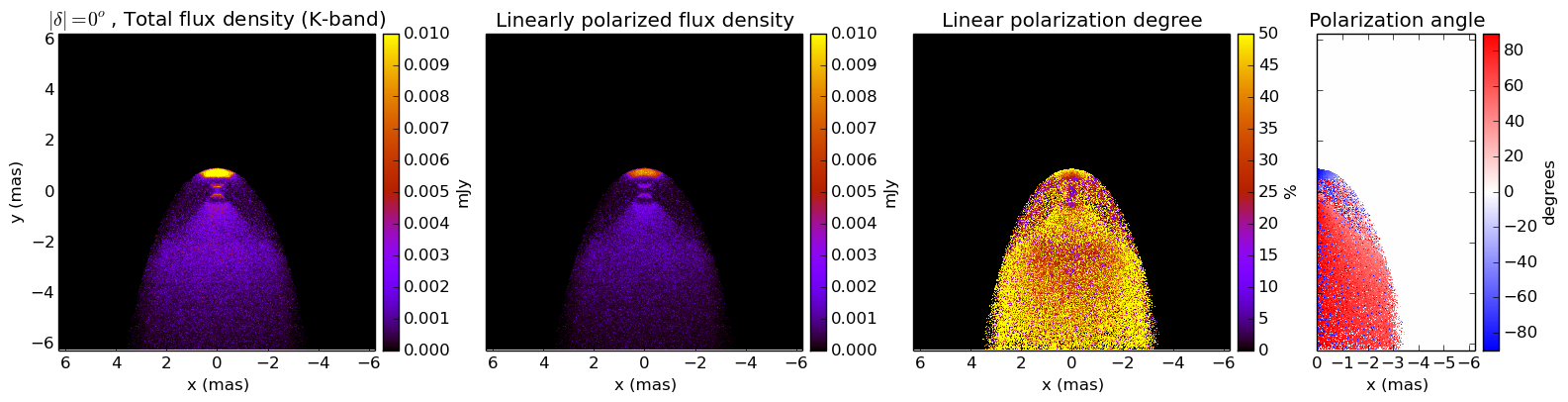}\\
   \includegraphics[width=\textwidth]{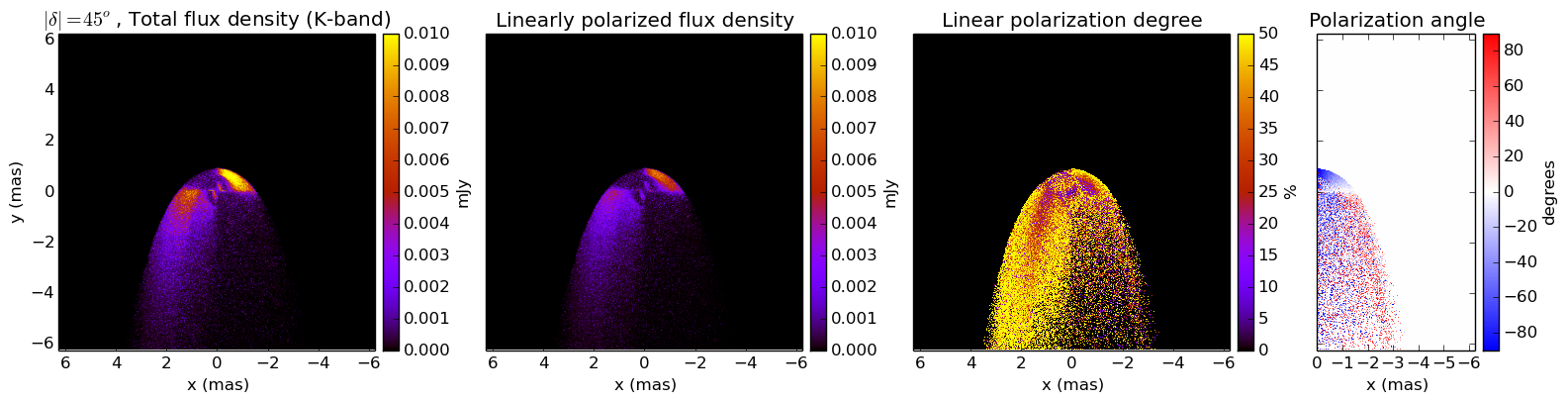}\\
   \includegraphics[width=\textwidth]{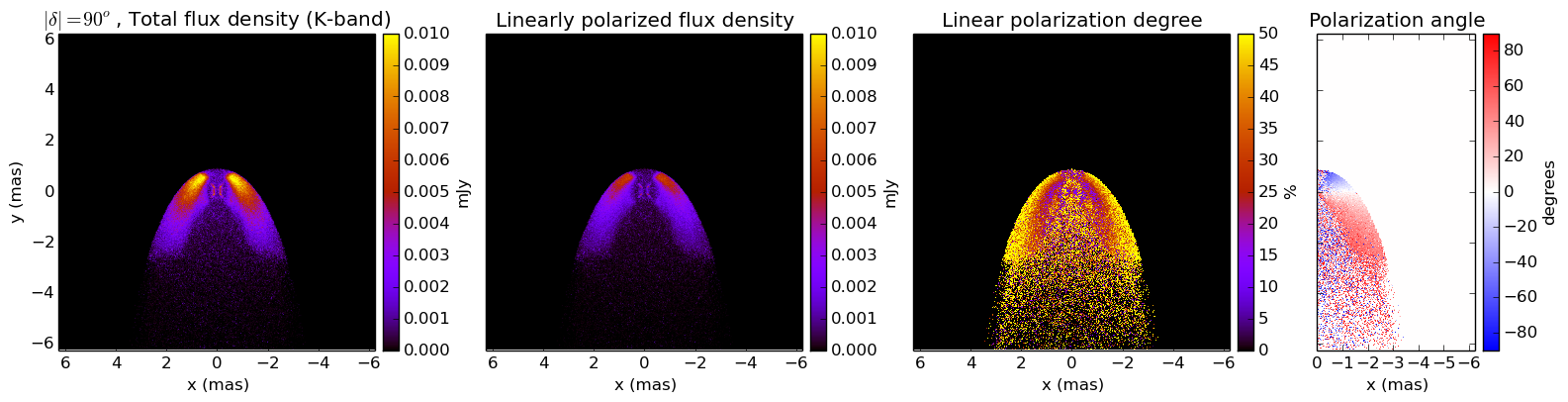}
 \end{tabular}
 \caption{Simulated images of the total flux density in $K_{\rm s}$ band (first panels from the left side), linearly polarized flux density (second panels from the left side), the distribution of the polarization degree (second panels from the right side), and the distribution of the polarization angle (first panels from the right side) for the different position angle of the bipolar outflow; $\delta=0^{\circ}$, $45^{\circ}$, and $90^{\circ}$ from the top to the bottom panels.}
 \label{img_pol_deg_angle_delta}
\end{figure*}

 The dependence of the total linear polarization degree in $K_{\rm s}$ band on the inclination is in Fig~\ref{fig_linpol_inc} (right panel). For radiative transfer calculations, we set the position angle of the bipolar cavities to $\delta=0^{\circ}$, i.e. aligned with the symmetry axis of the bow shock. The plot in Fig.~\ref{fig_linpol_inc} shows that a gradual increase in the bow-shock density (see also Fig.~\ref{fig_bowshock_density}) over the interval 2008--2012 leads to an increase in the total linear polarization degree over certain inclination ranges. A potential increase in the polarization degree is also discussed in \citet{Shahzaman2016}. It is, however, detected only for one epoch, $2012.0$ (see their Fig. 4 and Table 3), when the polarization degree reaches $p_{\rm L}\simeq 37.6\%$, so a progressive increase cannot be considered significant at this point. In addition, Figure~\ref{fig_linpol_inc} implies that the polarization degree can change when the inclination of the source geometry (bow shock), i.e. the viewing angle with respect to the observer, changes for different epochs. Indeed, this is the case for the DSO due to its fast motion along the elliptical orbit around Sgr~A*, see also the calculation of the viewing angle variation close to the pericentre passage presented in Appendix~\ref{appa1}. Furthermore, we perform the simulations for a bipolar outflow that is misaligned with the bow-shock symmetry axis. In Fig.~\ref{fig_linpol_inc}, two additional dependencies for $\delta=45^{\circ}$ and $\delta=90^{\circ}$ are calculated for the largest density (epoch 2012.0). Given our model set-up and an expected range of viewing angles (see the shaded area in Fig.~\ref{fig_linpol_inc}), the polarization degree of $\sim 30\%$ is better reproduced by the configuration, in which the position angle $\delta$ of the bipolar outflow is between $0^{\circ}$--$45^{\circ}$, i.e. more aligned towards the bow-shock symmetry axis. When the orientation of the bipolar outflow is perpendicular to the symmetry axis, $\delta=90^{\circ}$, the dependence of the polarization degree on the inclination is rather flat and stays below or around $20\%$.      

Qualitatively, all the polarization curves in Fig.~\ref{fig_dso_sed_inc} (right panel) have a peak that is close to $2.2\,{\rm \mu m}$  for the viewing angles around $90^{\circ}$, and the peak shifts towards shorter wavelengths for both smaller and larger viewing angles than $90^{\circ}$. The curves resemble the empirical Serkowski law, $p_{\rm L}=p_{\rm max}\exp{\{-K\ln^2{(\lambda_{\rm max}/\lambda)}\}}$ \citep{2015psps.book.....K}, where $\lambda_{\rm max}$ is the wavelength at which the polarization curve reaches the maximum $p_{\rm L}=p_{\rm max}$ and the coefficient $K$ determines the width of the curve. However, the Serkowski law is used to fit the linear polarization degree in the ISM that arises due to dichroic extinction, whereas the simulations presented here are performed for spherical grains, and without the implementation of the magnetic field, whose configuration at the studied distances from Sgr~A* is still highly uncertain and beyond the scope of this paper.

The variation of the position angle $\delta$ of the bipolar cavity leads to the change of the brightness distribution in the total as well as the linearly polarized light, and hence the change of the polarization angle $\Phi$. In \citet{Shahzaman2016} we fitted the dependency of the polarization angle on the position angle with the relation that is approximately equal to $\Phi \approx -(+)\delta + (-)90^{\circ}$ (see Fig. 9 in \citeauthor{Shahzaman2016}, \citeyear{Shahzaman2016}). This relation is also evident in the simulated images of the linear polarized light in Fig.~\ref{img_pol_deg_angle_delta}, where we change the position angle in $45^{\circ}$ steps from $0^{\circ}$ up to $90^{\circ}$ (from the top to the bottom panels in Fig.~\ref{img_pol_deg_angle_delta}). Most of the scattered, polarized light comes from the region where the bipolar cavities intersect the bow-shock shell. On the other hand, the minimum of the polarized emission is overlapping naturally with the optically thick dusty envelope that also hides the star at the centre. 

 \citet{Shahzaman2016} measured a variable polarization angle for four epochs, see their Fig. 4 (right panel). There are two possible mechanisms that can be employed to explain a variable polarization angle -- see also the left panel of Fig.~\ref{fig_dso_model} for the illustration:
\begin{itemize}
  \item[(i)] \textit{intrinsic changes in the star-envelope orientation:} these changes would be due to the torques induced by the massive black hole, which would lead to the precession of the circumstellar disc/bipolar outflows in case the disc is misaligned with respect to the orbital plane. The precession time-scale is longer than the orbital time-scale, $T_{\rm prec}>T_{\rm orb}$. On the other hand, the wobbling of the disc takes place on the time-scale shorter than one orbital period, approx. $T_{\rm wobble}\approx 1/2 T_{\rm orb}$ \citep{2000MNRAS.317..773B}, 
  \item[(ii)] \textit{external interaction of the star with the nuclear outflow/inflow:} such an interaction could change the viewing angle on the star-bow shock-bipolar outflow system, especially for the case when the outflow/inflow velocity is comparable to the orbital velocity of the star, which would significantly affect the relative velocity, $\mathbf{v_{\rm rel}}=\mathbf{v_{\rm star}}-\mathbf{v_{\rm{a}}}$, and hence also the orientation of the bow shock with respect to the observer, see the modelling by \citet{zajacek2016}.
\end{itemize}

\begin{figure}[tbh]
 \centering
 \includegraphics[width=0.475\textwidth]{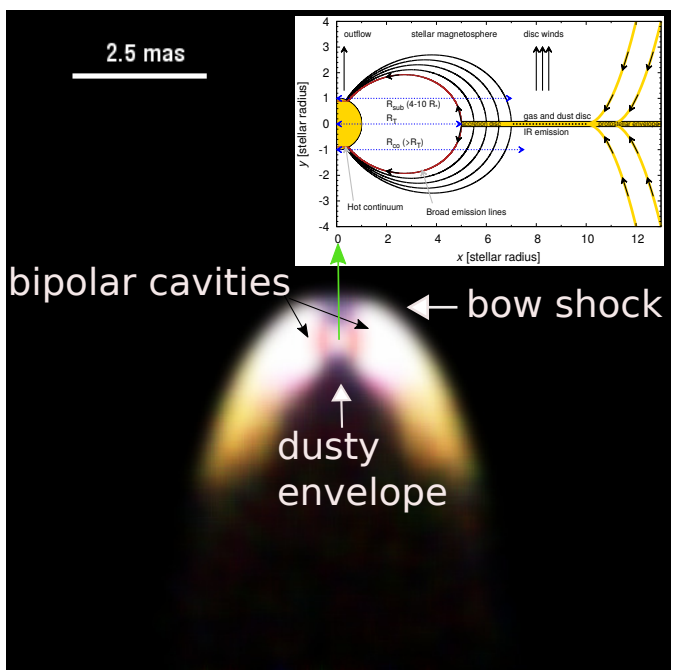}
 \caption{A simulated three-colour image of the source model of the Dusty S-cluster object (DSO/G2) for the position angle of the bipolar outflow $\delta=90^{\circ}$. Blue colour stands for $K_{\rm s}$ band, green colour for $L'$ band, and the red one for $M$ band emission. See \citet{Shahzaman2016} for an analogous composite image, but for the position angle $\delta=0^{\circ}$. The figure inset was adopted from \citet{2015wds..conf...27Z} and illustrates the magnetospheric accretion that takes place on the scale of several stellar radii and is possibly responsible for broad emission lines of the DSO.}
 \label{fig_rgb_image} 
\end{figure}

The simulated RGB image (Red colour - $L'$ band, Green colour - $M$ band, Blue colour - $K_{\rm s}$ band) of the source model of the DSO is in Fig.~\ref{fig_rgb_image} with the labels of the components. For the simulated image, we set the position angle $\delta$ to $90^{\circ}$, i.e. perpendicular to the symmetry axis of the bow shock (compare with the simulated image in Fig. 11 in \citeauthor{Shahzaman2016}, \citeyear{Shahzaman2016}, which was computed for $\delta=0^{\circ}$.) The inset in Fig.~\ref{fig_rgb_image} illustrates the magnetospheric accretion that was used to explain the origin of the broad Br$\gamma$ line of the DSO \citep{Valencia2015,2015wds..conf...27Z}.

Independently of the previous analysis, where the $K_{\rm s}$-band flux density is linearly polarized mainly due to dust photons scattering off spherical dust grains, a correlation was found between the linear polarization degree in $K_{\rm s}$ towards luminous stars embedded in molecular clouds and the optical depth $\tau_{\rm K}$, which can be fitted by a power law \citep{2015psps.book.....K},

\begin{equation}
  P_{\rm L,K}=2.2 \tau_{\rm K}^{0.75}\,(\%)\,.
  \label{eq_p_tau_correlation}
\end{equation} 
\citet{1992ApJ...389..602J} explain this correlation by a 50/50 mixture of the constant, i.e. perfectly aligned, component to the magnetic field with random components.  

When applied to the DSO with $P_{\rm DSO}\simeq 30\%$, we get $\tau_{\rm DSO} \approx 33$ as an estimate for the optical depth to the object along the line of sight, most of which can be attributed to the locally dense, optically thick envelope surrounding the stellar core. Similar values for the polarization degree and the optical depth in $K_{\rm s}$ band were found for the Becklin-Neugebauer object and OMC1-25 in the Orion star-forming region \citep{2015psps.book.....K}, which are both deeply embedded objects detected as prominent infrared excess sources. 

\subsubsection{Effect of surrounding stars on the polarized emission}

In the previous analysis, we assumed that the main source of photons that are absorbed and scattered off dust grains in the surrounding envelope is the star at the center. The question arises to what extent other stars in the S cluster contribute to the detected polarized emission. There are about $N_{\rm S}=30$ stars of mostly spectral type $B$ in the innermost arcsecond, $r_{\rm S}\approx 1''\simeq 0.04{\rm pc}$ \citep{Eckart2005}. Under the assumption of an approximately uniform distribution of stars in the sphere of radius $r_{\rm S}$, this gives the stellar number density in the S cluster $n_{\rm S}=N(4/3\pi r_{\rm S}^3)^{-1}\approx 10^5\,{\rm pc^{-3}}$. The average distance of any star from the DSO then is $\overline{D_{\rm S}}=(n_{\rm S})^{-1/3}\approx 0.02\,{\rm pc}=4.4\times 10^{3}\,{\rm AU}$. 

The ratio of total fluxes at the position where photons are scattered off grains is,

\begin{equation}
  \frac{F_{\rm DSO}}{F_{\rm S}}=\left(\frac{T_{\rm DSO}}{T_{\rm S}}\right)^{4} \left(\frac{R_{\rm DSO}}{R_{\rm S}}\right)^{2} \left(\frac{D_{\rm S}}{D_{\rm DSO}}\right)^{2}\,,
  \label{eq_flux_ratio}
\end{equation}    
where $T_{\rm DSO}$ and $T_{\rm S}$ are effective temperatures of the DSO and a typical S star, respectively. For the DSO, assuming it is a pre-main-sequence star, we take the previous value $T_{\rm DSO} \simeq 4200\,\rm{K}$. For a typical B0 star, the effective temperature is $T_{\rm S}\simeq 25\,000\,\rm{K}$. In terms of stellar radii, both the pre-main-sequence star and the B0 star have similar stellar radii of the order of  $R_{\rm DSO} \approx R_{\rm S}=10\,R_{\odot}$. Finally, distances of stellar sources from the scattering material are approximately $D_{\rm S}\approx \overline{D_{\rm S}} \simeq 4.4\times 10^{3}\,{\rm AU}$ and for the DSO star, $D_{\rm DSO} \approx 1\,{\rm AU}$. Plugging these estimated values into Eq.~\ref{eq_flux_ratio}, we get $F_{\rm DSO}/F_{\rm S} \approx 1.5 \times 10^{4}$, hence the contribution of other S stars is on average negligible in comparison with the central source.

An occasional close approach of an S star could contribute more. However, if these events were frequent, they should be reflected in a larger degree of variability of both the total and the polarized continuum emission. So far the DSO has appeared to be a rather stable source \citep{Shahzaman2016}.

\subsection{Possible non-thermal origin of the SED: NIR-``excess" sources as young neutron stars?}
\label{subsec_neutron_star}

In this subsection, we discuss the possibility that young neutron stars can in principle be detected in the central arcsecond of the Galactic centre and, under certain conditions, their characteristics would be similar to the Dusty S-cluster Object and other infrared excess sources, namely the positive spectral slope (larger flux density for longer wavelengths) and a significant polarized emission in NIR wavebands. 

As the first step, we check the energetics that would be required to produce flux densities comparable to the DSO. The flux density in $K_{\rm s}$ band is $F_{\nu}=0.23 \pm 0.02\,{\rm mJy}$ (see Table~\ref{tab_flux_density}). This leads to the overall $K_{\rm s}$ band luminosity of $L_{K}=\nu F_{\nu} 4 \pi r^2 = 0.23 \times 10^{-3} \times 10^{-23} \times 1.36 \times 10^{14} \times 4\pi\, (8000\, {\rm pc})^2\, {\rm erg\,s^{-1}}=2.4\times 10^{33}\,{\rm erg s^{-1}}$. This luminosity is of the same order of magnitude as the one found for PSR B0540-69 in the Large Magellanic Cloud \citep{2012A&A...544A.100M}. \citet{2012A&A...544A.100M} also show that the $K_{\rm s}$ band luminosity and the spin-down energy $\dot{E}$ are correlated, $L_{\rm K} \propto \dot{E}^{1.72 \pm 0.03}$, which implies that NIR emission is rotation-powered and associated with the magnetospheric origin and/or the neutron star wind termination shock. Using the correlation, $L_{\rm K} \approx 6.1 \times 10^{-33}\,\dot{E}^{1.72}$, we can estimate a spin-down power of the pulsar potentially associated with the DSO, $\dot{E}_{\rm DSO} \approx 1.4 \times 10^{38}\,{\rm erg\,s^{-1}}$, which is of the same order of magnitude as the Crab pulsar, J0534+2200 \citep{2005AJ....129.1993M}.  

Hence, the neutron star associated with the DSO and other NIR-excess sources would have to be a rather young, Crab-like pulsar wind nebula (PWN) with the characteristic age of $\tau=P/2\dot{P}\approx 10^{3}\,{\rm yr}$, where $P$ is a pulsar period and $\dot{P}$ is the period derivative or spin-down rate. The origin would be young, massive OB stars having an age of a few millions observed in the central parsec \citep{2009A&A...499..483B}. The power-law slope inferred for the DSO, see Eq.~\eqref{eq_nonthermal_fit}, is qualitatively consistent with the observations of neutron stars in near-infrared bands \citep{2012A&A...544A.100M,2013A&A...554A.120Z,2016MNRAS.455.1746Z}, however, it appears to be steeper than for observed pulsars \citep{2012A&A...544A.100M}, which have the mean spectral index $\alpha \approx 0.7$. 

The SED alone does not give any convincing argument for the neutron star hypothesis and the dust-enshrouded star is thought to be a more natural scenario. On the other hand, the detection of linearly polarized emission in $K_{\rm s}$ band and a high polarization degree of $\sim 30\%$ \citep{Shahzaman2016} imply that the DSO may indeed be a peculiar source in the S cluster and the neutron star model can naturally explain the polarized emission via the synchrotron mechanism in the dipole magnetic field, see Eq.~\eqref{eq_neutronstar_pol_deg}. For instance, the infrared imaging and polarimatric observations of the pulsar wind nebula SNR G21.5-0.9 \citep{2012A&A...542A..12Z} indeed show a high degree of linear polarization in $K_{\rm s}$ band, $P_{\rm L}\simeq 0.47$ and a comparably high polarization degree is expected for other pulsar wind nebulae.

Young, rotation-powered neutron stars are expected to have radio and X-ray counterparts whose luminosities are proportional to the $\dot{E}_{\rm rad}=-\dot{E}_{\rm spin-down}=4\pi^2 I \dot{P}/P^3$, where $I$ is the moment of inertia of the neutron star, $I\approx 10^{45}\,{\rm g\,cm^2}$. In general, there seems to be a trend of increasing radiative efficiency $\eta_{\rm f}$ towards shorter wavelengths, $\eta_{\rm f}\equiv L_{\rm f}/\dot{E}_{\rm rad}$, where $f$ is the spectral domain \citep{2012hpa..book.....L}. 

In the X-ray domain the scatter of efficiencies is relatively small, and approximately equal to $\eta_{\rm X}\approx 10^{-4}$ \citep{2007ApJ...670..655K}, which in case of the pulsar and its wind nebula associated with the DSO gives $L_{\rm X}=\eta_{\rm X} \dot{E}_{\rm DSO}\approx 10^{34}\,{\rm erg\,s^{-1}}$, which is, given the uncertainties, comparable to the quiescent X-ray emission of Sgr~A*, $L_{\rm X,Sgr~A*}\approx 10^{33}\,{\rm erg\,s^{-1}}$ \citep{2014ARA&A..52..529Y}, associated with the thermal bremsstrahlung process in the hot plasma surrounding Sgr~A*.

Towards the radio domain, the radiative efficiency for rotation-powered pulsars becomes smaller and the scatter is larger, $\eta_{\rm R}=10^{-8}$--$10^{-5}$, which leads to the values for the pulsar associated with the DSO, $L_{\rm R}=\eta_{\rm R} \dot{E}_{\rm DSO}\approx 10^{30}\,{\rm erg\,s^{-1}}$--$10^{33}\,{\rm erg\,s^{-1}}$, which is smaller or comparable to the luminosity of Sgr~A* in this domain \citep{2014ARA&A..52..529Y}. In some cases, pulsars are only detected at higher energies and appear to be radio-quiet (e.g. Geminga pulsar; \citeauthor{2003ASPC..302..145C}, \citeyear{2003ASPC..302..145C}) under the sensitivity constraints of radio surveys, which either implies that the radio beam is narrow and not directed at the Earth or that radiative efficiencies in the radio domain for some young pulsars are smaller in comparison with X-ray and $\gamma$-ray domains.  

\begin{figure}[tbh]
  \centering
  \includegraphics[width=0.5\textwidth]{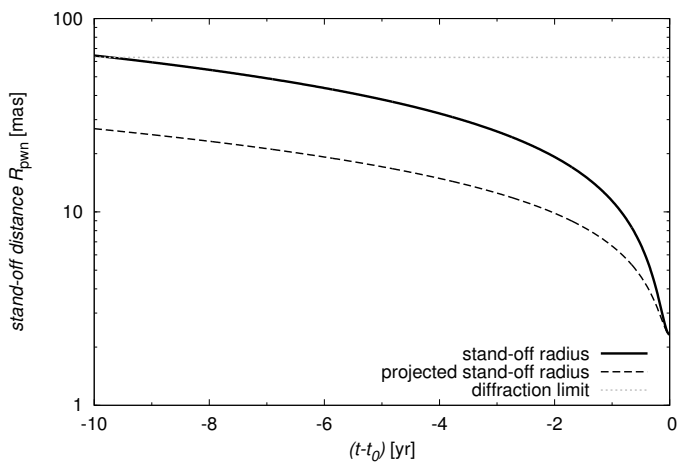}
  \caption{The evolution of the stand-off distance (in mas) of the pulsar wind nebula as a function of time (in years before the pericentre). The total  luminosity is set to $\dot{E}_{\rm rad}=1.4 \times 10^{38}\,{\rm erg\,s^{-1}}$, the density of the ambient medium is computed according to Eq.~\eqref{eq_density}, and the orbital velocity is calculated using the orbital elements of the DSO as inferred by \citet{Valencia2015} using Br$\gamma$ emission of the DSO.}
  \label{fig_standoff_pwn}
\end{figure}

So far no clear X-ray and radio counterparts of the DSO were detected \citep{bower2015,mossoux2016,borkar2016}. However, given the rough estimates above, a high background flux towards Sgr~A*, and low angular resolution in the X-ray domain, even young pulsars of a Crab type can be beyond the sensitivity limits of current X-ray and radio instruments. Therefore, weak, infrared excess sources similar to the DSO could be candidates for pulsars and deserve follow-up monitoring with upcoming, high-sensitivity facilities (E-ELT, Square Kilometer Array -- SKA). 

Given the large orbital velocities of stars and potential stellar remnants in the S cluster with respect to the ambient medium, pulsar wind nebulae are expected to be non-spherical and elongated in the direction of the motion. The length-scale of the termination-shock of the pulsar wind is given by the pressure balance as expressed by Eq.~\eqref{eq_standoff}, where the wind pressure term $p_{\rm w}=\dot{m}_{\rm w} v_{\rm w}/(4\pi r^2)$ is replaced by $p_{\rm pwn}=\dot{E}_{\rm rad}/(4 \pi c r^2)$, which leads to the stand-off distance

\begin{equation}
 R_{\rm PWN}=\left(\frac{\dot{E}_{\rm rad}}{4 \pi c \rho_{\rm a} v_{\rm rel}^2}\right)^{1/2}\,,
 \label{eq_rpwn}
\end{equation}
where we neglected the thermal term. The relative velocity with respect to the ambient medium may be approximated by the orbital velocity close to the pericentre of the orbit, $v_{\rm rel} \simeq v_{\star}$. Taking $\dot{E}_{\rm rad}=1.4 \times 10^{38}\,{\rm erg\,s^{-1}}$ as estimated for the pulsar potentially associated with the DSO, we calculate the evolution of the stand-off distance according to Eq.~\eqref{eq_rpwn} during the period of $10$ years before the pericentre passage, see Fig.~\ref{fig_standoff_pwn}. We see that even for a relatively young PWN with the spin-down energy of the same order as the Crab pulsar, the stand-off distance is comparable or smaller than the diffraction limit of $63\,{\rm mas}$, $R_{\rm PWN} \leq \theta_{\rm min}$, i.e. the PWN would effectively appear as a point source, mainly due to a large relative velocity close to Sgr~A*.

A detailed analysis of the generation of Br$\gamma$ line in the pulsar wind nebula is beyond the scope of this paper. We just note that the source of Br$\gamma$ could be the collisional excitation of hydrogen atoms prior to their ionization in the bow-shock layer, as was similarly suggested by \citet{Scoville2013} for a bow shock associated with a T Tauri star. Specifically, there are several pulsar bow shocks that exhibit $H\alpha$ emission \citep{1996ASPC..105..393C,2014ApJ...784..154B}, and hence Br$\gamma$ line could be produced by the same process. Quantitatively, for three pulsars at kiloparsec distances -- B0740-28, B1957+20, B2224+65 (see Table 2 in \citeauthor{2002ApJ...575..407C}, \citeyear{2002ApJ...575..407C}) -- the corresponding $H\alpha$ luminosities are in the range of $L_{{\rm H}\alpha}\simeq 6.5\times 10^{28}$--$1.6 \times 10^{30}\,{\rm erg\,s^{-1}}$, i.e. $L_{{\rm H}\alpha}\simeq 2 \times 10^{-5}$--$4\times 10^{-4}\,L_{\odot}$, which is at least an order of magnitude less than the Br$\gamma$ luminosity of the DSO \citep{Gillessen2012,Valencia2015}. In general, \citet{2002ApJ...575..407C} derive a scaling relation for H$\alpha$ luminosity of pulsars, $L_{{\rm H}\alpha} \propto 4\pi X\dot{E}_{\rm rad}v_{\star}$, where $X$ is a fraction of neutral hydrogen. Analogous dependencies are expected for Br$\gamma$ luminosity, under the assumption that the line is produced by the collisional excitation in the pulsar bow shock.    
 
In addition, natal kick velocities during supernova explosions in the clockwise stellar disk could, under certain configuration, lead to the formation of highly-eccentric orbits similar to that of the DSO; see the analysis presented in Paper II.  

In summary, given the properties of the DSO and that of other NIR-excess sources, one cannot a priori exclude the PWN hypothesis as an explanation of the phenomenon.

\section{Discussion}
\label{sec_discussion}

The focus of this paper was on the detailed analysis of the total and polarized continuum emission of the DSO in the NIR domain as explained by an enshrouded pre-main-sequence star model. A detailed analysis of other phenomena associated with the DSO, i.e. a tentative association with a larger tail/streamer, emission in other domains, and the analysis of the line emission are beyond the scope of this study. However, let us briefly comment on some of these aspects to put the continuum analysis in a larger context. These and other aspects of NIR-excess sources in the Galactic centre will also be examined in detail in upcoming papers of this series.

\subsection{DSO/G2 and G1 as parts of a larger gaseous streamer?}
 It was claimed previously that DSO/G2 is accompanied by an elongated, low-surface-brightness tail that is clearly visible in the position-velocity diagrams \citep{Gillessen2012,Gillessen2013a,Gillessen2013b,Pfuhl2015}. A dynamical connection of the DSO/G2 to the tail would favour the ``gas-stream-like” scheme: according to this scenario, DSO would be a clump of a more extended filamentary structure. However, this clump could still be compact and associated with a star; see also \citet{Ballone2016} for the hydrodynamical model of the ``G2$+$G2t'' complex. Indeed, the numerical models of the formation of young stars close to the supermassive black hole suggest that the star formation takes place in infalling gaseous clumps that undergo fragmentation and stretching \citep{Jalali2014}. The occurence of denser clumps -- protostars -- embedded in a larger streamer is therefore expected in the earliest stages of the star formation in the strong gravity regime, where tidal forces cannot be neglected.   
 
 However, apart from the arguments above, the connection of the DSO to a larger filamentary structure is still disputed observationally, mainly because of a spatial and kinematical offset from the DSO \citep{eckart2014a, Meyer2014a, Phifer2013}; see also Peissker et al. (to be submitted). It is plausible that the tail is rather a back/foreground feature associated with the minispiral \citep{eckart2014a}. The denser filament in the south-east direction could be a result of the interaction of the northern and the eastern minispiral arms whose dynamics is consistent with two Keplerian bundles (see Fig. 21 in \citeauthor{2009ApJ...699..186Z}, \citeyear{2009ApJ...699..186Z}, and Fig. 10 in \citeauthor{2000NewA....4..581V}, \citeyear{2000NewA....4..581V}). \citet{Meyer2014a} also claim that while the DSO and the filament have similar radial velocities, there is a significant spatial offset, which is not visible in position-velocity plots of \citet{Gillessen2012,Gillessen2013a,Gillessen2013b} and \citet{Pfuhl2015}, who use an artificial curved slit of a finite width to extract radial velocities, which masks spatial offsets. This also means that the filament can be an artefact as the result of the masking process in the data reduction procedure, i.e. smaller line-emitting sources that have an offset are grouped together. 
 
 The association of the PWN to a larger filament is also plausible, since it could be a remnant of the material from the supernova explosion, e.g. in the clockwise disc of massive OB stars. In this sense, the fast-moving pulsar can ``catch up" with the expanding shell, as also seems to be the case of Sgr~A~East remnant and the ``Cannonball" pulsar \citep{2013ApJ...777..146Z}. 
 
 \citet{Pfuhl2015} also propose a possible dynamical connection between the DSO/G2 object and G1 clump, i.e. both features could be parts of the same streamer and dynamically related. If this was the case, it could be used for constraining the properties of the accretion flow around Sgr~A* \citep{2016MNRAS.455.2187M, 2017MNRAS.465.2310M}, based on the amount of dragging, which is only prominent for core-less clumps. However, the association of these two, reddened sources is still under investigation and should be confirmed/excluded based on continuing monitoring -- even if the inclination of both orbits is comparable, the orientation of the orbits in space can be different. Also, G1 object survived the pericentre passage, so it also seems less likely that it is a pure gas cloud \citep{2015AAS...22510207S}. 
 
 In summary, even if the DSO was associated with or moving through a larger gaseous structure, it would not rule out a young-star or PWN hypothesis presented in this study.
 
\subsection{Core-less cloud models vs. stellar models of the DSO}  
Following the roadmap in Fig.~\ref{fig_roadmap}, we will look at a few observables that distinguish the core-less cloud and stellar interpretation of the DSO.

 A dynamical comparison of the evolution of a core-less cloud and a stellar model using a simplified test particle model was done in \citet{Zajacek2014}. The basic result for an assumed density profile of the ambient medium was that a core-less cloud starts progressively deviating from the Keplerian orbit soon after the pericentre, where the velocity of the cloud as well as the density of the medium are the highest and therefore the non-gravitational acceleration due to a hydrodynamical drag is the largest as well; see also \citet{2016MNRAS.455.2187M} and \citet{2017MNRAS.465.2310M}. The star is naturally much less sensitive to the ambient medium because of the smaller ratio of the cross-section to the mass, and follows the original trajectory. Up to now, the DSO does not deviate within uncertainties from the inferred highly-eccentric orbit (Peissker et al., in prep.), which may indicate a stellar nature. However, it could also be argued that the hydrodynamical drag is smaller due to a smaller ambient density than usually assumed in the models of RIAFs. 
 
 A stronger argument in favour of a stellar model is the compactness of the source as discussed in detail in Section~\ref{sec_compactness}. In $K_{\rm s}$ and $L'$ band continuum, the DSO is fully consistent with a point source \citep{eckart2014a, Witzel2014,Shahzaman2016}. Concerning the analysis of Br$\gamma$ line emission, there are contradictory observational results. While \citet{Valencia2015} detect a single-peak Br$\gamma$ emission at each epoch, both shortly before and after the peribothron passage (see their Fig. 8), which is consistent with a stellar model, \citet{Pfuhl2015}, on the other hand, detect both a red-shifted peak and a blue-shifted peak for the same epoch (see their Fig. 15), which indicates the effect of stretching of the gas along the orbit. Despite this observational contradiction, it is difficult to reconcile an apparent compact continuum emission (tracing dust) and a stretched Br$\gamma$ emission (tracing gas) with the gas cloud model, since such a cloud would be expected to consist of a well-mixed gas-dust components. This was interpreted by \citet{Witzel2014} in terms of a binary merger model, which is accompanied by an expanding dusty outflow forming an optically thick photosphere \citep{Pejcha2014}. Since binary mergers contract on a Kelvin-Helmholtz time-scale and are surrounded by a dusty envelope, they share several characteristics with pre-main-sequence stars. Therefore, our dust-enshrouded star model is applicable to this scheme. 
 
 In \citet{Schartmann2015}, a compact cloud model is presented, where the authors claim that the cloud follows the Keplerian orbit and stays rather compact. However, for the epochs 2005--2015 the full width at half maximum of the normalized Br$\gamma$ emission in their model is clearly above the diffraction limit of $63\,{\rm mas}$ and reaches as much as $115\,{\rm mas}$ at the pericentre. Therefore, the cloud should have been clearly resolved at all epochs of the monitoring. In their model, the cloud also undergoes tidal stretching and disruption, with an expected enhancement in accretion as early as $2014.6$ (see their Fig. 1), which is not consistent with the monitoring of Sgr~A*  \citep{borkar2016, bower2015, mossoux2016}. This indicates a rather unchanged luminosity of the quiescent state as well as the overall flare statistics.

\subsection{Contribution of the DSO passage to the variability of Sgr~A*?}

Concerning the passage of the DSO close to the Galactic centre, there are two possibilities of how the source could have affected the radiative properties of Sgr~A*:
\begin{itemize}
  \item[(i)] non-thermal emission associated with the shock front,
  \item[(ii)] increase in the flaring activity of Sgr~A* due to the changing accretion-flow properties close to Sgr~A*.
\end{itemize}  

There have been several multi-wavelength observational campaigns that aimed at capturing the signs of the DSO/Sgr~A* interaction. The X-ray flux density of Sgr~A* was monitored by \textit{Chandra} and no increase in the quiescent X-ray emission was detected during 2013-2014 campaign \citep{2014ATel.6242....1H}, which collected $\sim 900\,{\rm ks}$ of data. Also, no change in the radio quiescent emission was detected \citep{bower2015,borkar2016}. This implies that the shock associated with the supersonic motion of the DSO was far less strong and bright than the predicted values \citep{2012ApJ...757L..20N} that exceeded the quiescent level of Sgr~A*. It means that either the actual source has a smaller cross-section than the one assumed for synchrotron calculations, $\sigma\lesssim 2\times 10^{29}\,{\rm cm^2}$ that corresponds to the diameter of $D\lesssim 34\,{\rm AU}$ \citep{bower2015}, or that the ambient medium is rather rarefied even in comparison with the original RIAF models.    

Concerning the flaring activity of Sgr~A*, there is no statistically significant enhancement detected in both X-ray and infrared bands \citep{Valencia2015, mossoux2016}. Based on the analysis of $\sim 15\,{\rm yr}$ of \textit{XMM}, \textit{Chandra}, and \textit{Swift} data, \citet{ponti2015} discuss an increase in the rate of bright X-ray flares. However, an alternative and plausible explanation is an effect of the X-ray flare clustering \citep{2017IAUS..322....1H} rather than the actual increase in the flaring rate, so the overall statistics does not seem to be affected by the DSO passage \citep{mossoux2016}.

Nevertheless, it is of interest to estimate the non-thermal synchrotron or potential X-ray bremsstrahlung flux for the stellar model of the DSO. These estimates may also be scaled for different stellar parameters when other stars will pass in the vicinity of Sgr~A* and future facilities with the larger sensitivity may detect the corresponding emission.

\subsubsection{Bow-shock synchrotron emission}

\begin{figure*}[tbh]
  \centering
  \begin{tabular}{cc}
    \includegraphics[width=0.5\textwidth]{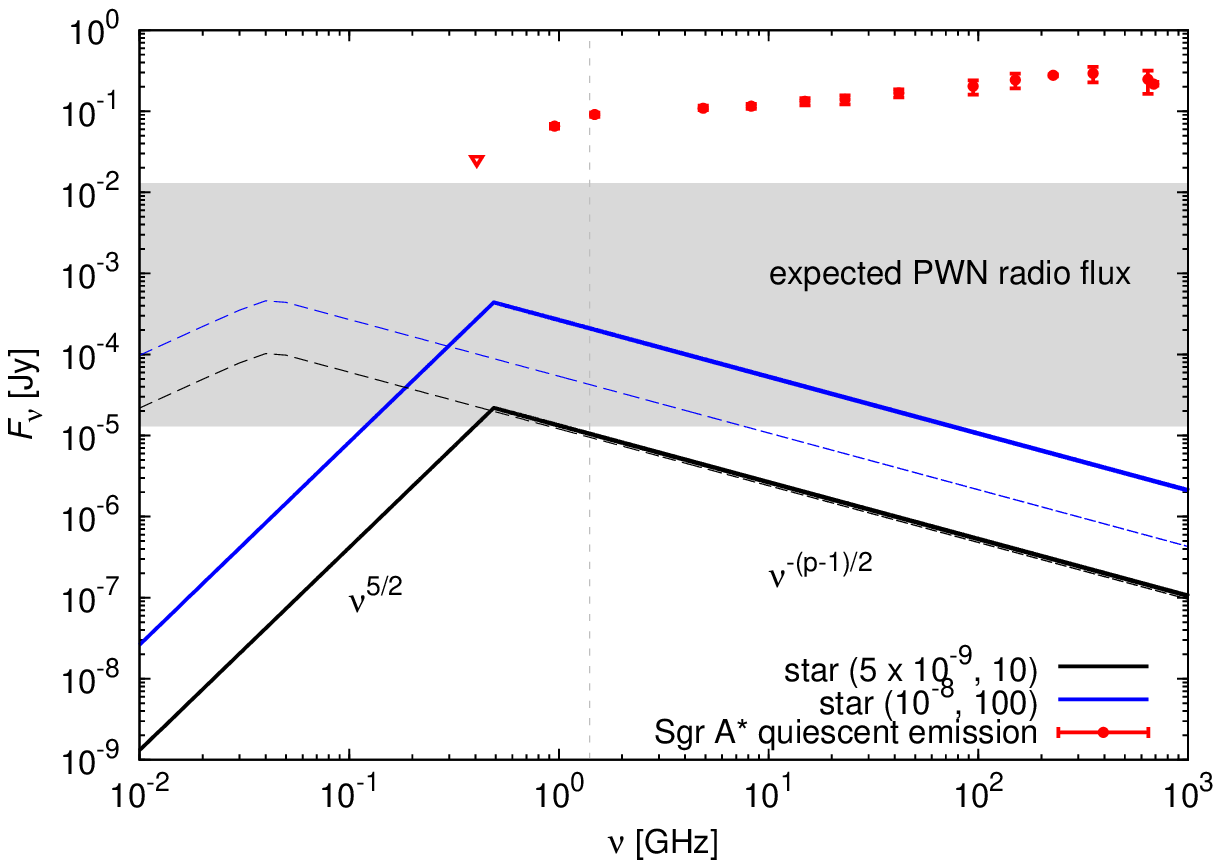} & \includegraphics[width=0.5\textwidth]{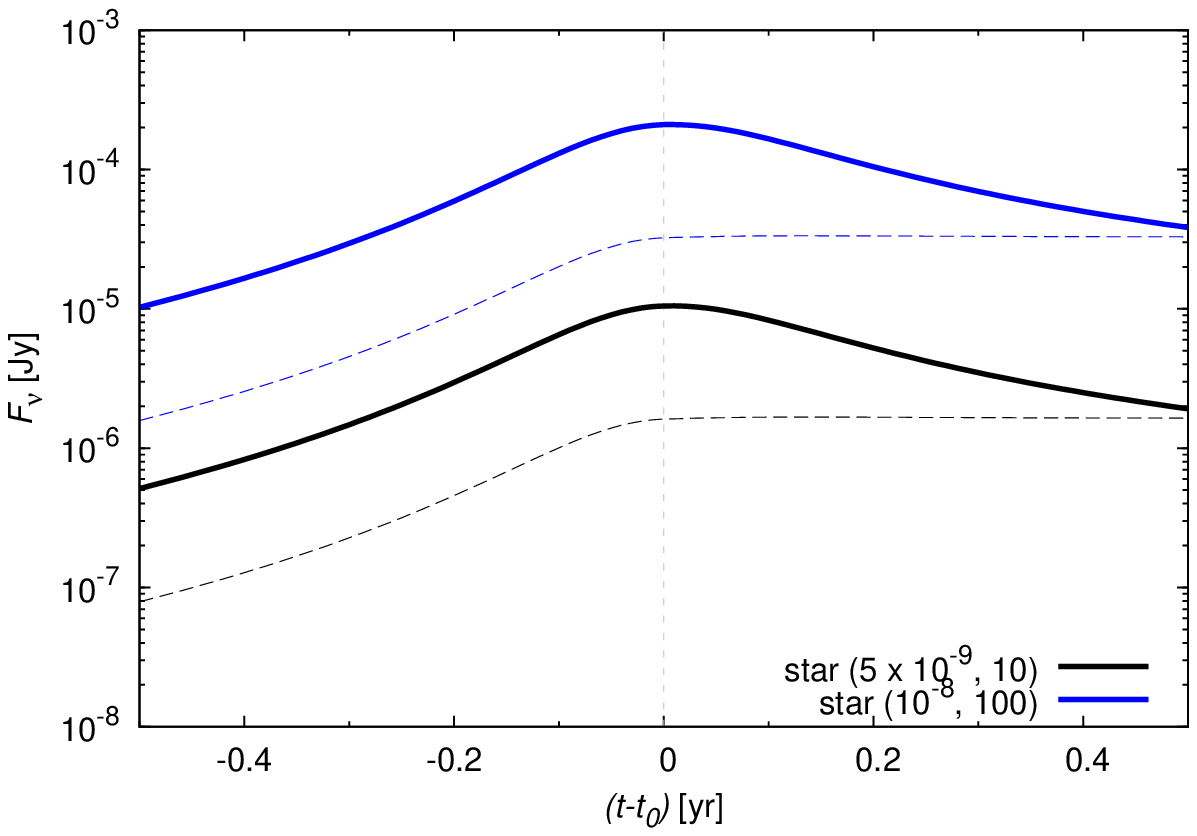}
  \end{tabular}
  \caption{\textbf{Left panel:} Calculated synchrotron spectrum for electrons accelerated in the bow shock. Spectra were calculated for two sets of the mass-loss rate and the terminal wind velocity that are within the range of values expected from the stellar model: $\dot{m}_{\rm w}=5\times 10^{-9}\,M_{\odot}\rm{yr^{-1}}$, $v_{\rm w}=10\,{\rm km\,s^{-1}}$ and $\dot{m}_{\rm w}=10^{-8}\,M_{\odot}\rm{yr^{-1}}$, $v_{\rm w}=100\,{\rm km\,s^{-1}}$. Solid lines correspond to the \textit{plowing} synchrotron model and the dashed lines to the \textit{local} model. The energy spectrum of accelerated electrons has a power-law index of $p=2.4$ in accordance with particle-in-cell simulations of \citet{Sadowski2013}. The red points are observationally determined values adopted from \citet{1976MNRAS.177..319D}, \citet{2000A&A...362..113F}, \citet{2003ApJ...586L..29Z}, and \citet{2008ApJ...682..373M}. The shaded region represents an estimate of the radio flux density of the PWN. \textbf{Right panel:} A synthetic light curve calculated at $1.4\,{\rm GHz}$ considering for both the plowing (solid lines) and the local model (dashed lines). Two sets of stellar parameters were considered as in the left panel.}
  \label{fig_synchrotron} 
\end{figure*}

The source of the non-thermal synchrotron emission of the DSO are the electrons accelerated in the bow shock. We apply the theory and the model of \citet{crumley2013} and \citet{Sadowski2013} \citep[based on][]{Rybicki1979} to calculate synchrotron spectra and the light curve of the source. In the left panel of  Fig.~\ref{fig_synchrotron}, the spectra are calculated for the epoch of the pericentre passage, when the flux density is the largest across all frequencies. The right panel of Fig.~\ref{fig_synchrotron} displays the light curve computed for the frequency of $1.4\,{\rm GHz}$. For both the spectrum and the light curve, two limiting cases are taken into account:
\begin{itemize}
\item[(i)] \textit{``plowing"} model that assumes that all electrons accelerated in the bow shock radiate in the shock-enhanced magnetic field,  
\item[(ii)] \textit{``local"} model that assumes that the electrons accelerated in the shock leave it and radiate in the local, ambient magnetic field.
\end{itemize}

Two sets of stellar parameters were considered: the mass-loss rate of $\dot{m}_{\rm w}=5\times 10^{-9}\,M_{\odot}\,{\rm yr^{-1}}$ and the wind velocity of $v_{\rm w}=10\,{\rm km\,s^{-1}}$ and $\dot{m}_{\rm w}=10^{-8}\,M_{\odot}\,{\rm yr^{-1}}$, $v_{\rm w}=10\,{\rm km\,s^{-1}}$, which are expected for a class 1 pre-main-sequence source \citep{2006ApJ...646..319E} with the empirical relation  $\dot{M}_{\rm w}\approx 0.001-0.1\,\dot{M}_{\rm acc}$ between the accretion rate and the outflow rate. For both parameter sets, the peak flux densities are at least two orders of magnitude smaller than the quiescent radio flux density of Sgr~A*, i.e. the stellar model can naturally explain the non-detection of any enhancement in the flux density due to a small bow-shock size at the pericentre. In addition, the values are comparable to the radio flux density of the pulsar wind nebula, taking into account the radio efficiencies of $\eta_{\rm R}\approx 10^{-8}-10^{-5}$ and the spin-down luminosity of $\dot{E}_{\rm rad}=1.4 \times 10^{38}\,{\rm erg\,s^{-1}}$.

In summary, both stellar models -- a dust-enshrouded pre-main-sequence star and a PWN -- are expected to have radio flux densities below the quiescent level of Sgr~A*, implying only a marginal possibility of detection of any change in the radio flux of Sgr~A* and its close vicinity.

\subsubsection{X-ray bremsstrahlung}   

\begin{figure}[tbh]
   \centering
   \includegraphics[width=0.5\textwidth]{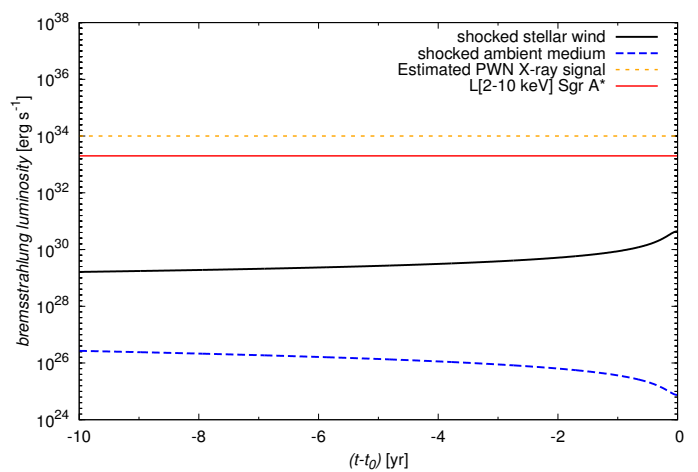}
   \caption{Temporal evolution of the X-ray bremsstrahlung signal from two components: shocked wind (black solid line) and shocked ambient medium (blue dashed line). Two vertical lines represent the quiescent emission of Sgr~A*, $L[2-10\,{\rm keV}]\sim 2\times 10^{33}\,{\rm erg\,s^{-1}}$ \citep{2017IAUS..322....1H}, and the estimated PWN signal, assuming the conversion factor of $\eta_{\rm X}=10^{-4}$ between the spin-down luminosity and the X-ray luminosity. The stellar parameters were set to $\dot{m}_{\rm w}=10^{-8}\,M_{\odot}{\rm yr^{-1}}$ and $v_{\rm w}=100\,{\rm km\,s^{-1}}$.}
   \label{img_brems}
\end{figure}

Due to an expected supersonic motion of the DSO close to the pericentre \citep{zajacek2016}, two layers of shocked stellar wind and ambient gas should form and subsequently mix. The shocked gas is expected to cool down by the thermal bremsstrahlung process \citep{2016MNRAS.459.2420C}. Here we use order-of-magnitude estimates to calculate the bremsstrahlung luminosity. 

First, the ionized medium in the central parsec of the Galactic centre can in principle cause losses to the bremsstrahlung due to the Thomson scattering. We can estimate the optical depth of the scattering using $\tau_{\rm T} \sim \sigma_{\rm T} n_{\rm e} L$, where $\sigma_{\rm T}=6.65 \times 10^{-25}\,{\rm cm^2}$ is the cross-section for the Thomson scattering, $n_{\rm e}$ is the electron number density, and $L$ is the length of the part of the line of sight that crosses the hot gas. For an estimate, we take the number density of electrons at the Bondi radius, $R_{\rm B}\approx 0.16 (T_{\rm a}/10^7\,{\rm K})^{-1}\,{\rm pc}$, $n_{\rm e}(R_{\rm B})=130\,{\rm cm^{-3}}$ \citep{2003ApJ...591..891B} and the line-of-sight length is set to the Bondi radius, $L\simeq R_{\rm B}$. Finally, the estimate gives $\tau_{\rm T} \sim 4.3 \times 10^{-5} \ll 1$ and therefore the Thomson scattering losses can be neglected. 

We take the volume emission coefficient per unit frequency band and integrate it over the frequency interval of the X-ray band, $(\nu_1, \nu_2)=(0.04,0.8)\times 10^{18}\,{\rm Hz}$. For the volume emission coefficient $\epsilon_{\nu}$ holds \citep{1978afcp.book.....L,Rybicki1979},

\begin{align}
  \epsilon_{\nu} \mathrm{d}\nu \approx 5.4 \times 10^{-39} Z^2 T_{\rm sh}^{-1/2} n_{\rm i} n_{\rm e}\, g(\nu, T_{\rm sh})\,\times\,\notag\\
    \times \exp{(-h\nu/kT_{\rm sh})} \mathrm{d}\nu\,\,{\rm erg\,s^{-1}\,cm^{-3}\,Hz^{-1}}\,,\label{eq_bremsstrahlung_coefficient}
\end{align} 
where $Z$ is the proton number of participating ions, $n_{\rm i}$  and $n_{\rm e}$ are number densities of ions and electrons, respectively, and $T_{\rm sh}$ is the temperature of the shocked gas. The temperature of the shocked ambient medium for the supersonic motion of the DSO may be estimates as $T_{\rm sh}=1.38 \times 10^5 (v_{\star}/100\,{\rm km\,s^{-1}})^2\,{\rm K}$ for an ionized medium \citep{1980ARA&A..18..219M}, which yields $T_{\rm sh} \sim 10^{8}-10^9\,{\rm K}$ for relative velocities of $v_{\star}\sim 10^3\,{\rm km\,s^{-1}}$ close to the pericentre. The shocked stellar wind is colder by about three orders of magnitude \citep{Scoville2013}, but denser, so it is likely to contribute more to the overall bremsstrahlung. 

The Gaunt factor $g(\nu, T_{\rm sh})$ may be approximated in the given range ($T_{\rm sh}=10^5-10^8\,{\rm K}$, $\nu=10^{18}\,{\rm Hz}$) via \citep{1978afcp.book.....L}
\begin{equation}
  g(\nu, T_{\rm sh})\sim \sqrt{3}/\pi \log{(4.7 \times 10^{10} T_{\rm sh}/\nu)}
  \label{eq_gaunt_factor}
\end{equation}  
which varies only a little for the considered frequency range. According to \citet{eckart1983}, we set $g(\nu, T_{\rm sh})=1.5$. Using Eq.~\eqref{eq_bremsstrahlung_coefficient}, the bremsstrahlung luminosity can be expressed as,

\begin{align}
  L_{\rm brems}\sim 5.4\times 10^{-39}\times 4\pi \times 1.5 V_{\rm sh} Z^2 T_{\rm sh}^{-1/2} n_{\rm i} n_{\rm e} \times \,\notag\\
   \times \int_{\nu_1}^{\nu_2}\exp{(-h\nu/kT_{\rm sh})}\mathrm{d}\nu\,{\rm erg\,s^{-1}}\,.\label{eq_brems_luminosity}
\end{align}

The integral $\int \exp{(-h\nu/kT_{\rm sh})} \mathrm{d}\nu$ yields $\sim 0.63 \times 10^{10}\,T_{\rm sh}$ \citep{eckart1983}. Furthermore, we express the temperature of shocked gas in $T_{8}\equiv (T_{\rm sh}/10^{8}\,{\rm K})$ and assume that the shocked gas consists fully of ionized hydrogen and helium, so one can replace the term $Z^2 n_{\rm i} n_{\rm e}=1.55 n_{\rm e}^2$. We approximate the volume of the shocked gas by a spherical half-shell with the volume of $V_{\rm sh}\sim 2\pi R_{0}^3 f_{\rm HR}$, where $f_{\rm HR}$ is the ratio of the bow-shock thickness to the stand-off radius, Eq.~\eqref{eq_standoff}. Finally, the bremsstrahlung luminosity becomes

\begin{equation}
  L_{\rm brems} \approx 6.2 \times 10^{-23} R_{0}^3(t) f_{\rm HR}(t) n_{\rm e}^2 T_8^{1/2}\,{\rm erg\,s^{-1}}\,,
  \label{eq_brems_luminosity}
\end{equation}
where both the stand-off radius $R_0(t)$ and the ratio $f_{\rm HR}(t)$ depend on the orbital phase (time) of the DSO \citep{zajacek2016,2016MNRAS.459.2420C}.

We calculate the bremsstrahlung luminosity using Eq.~\eqref{eq_brems_luminosity} for both shocked layers, ambient and stellar wind layer. The density of the shocked stellar wind is computed using Eq.~\eqref{eq_bowshock_density} and we take the temperature of $T_{\rm sh} \simeq 10^5\,{\rm K}$. The shocked ambient medium is denser than the ambient profile, Eq.~\eqref{eq_density}, by about a factor of four and its temperature reaches $T_{\rm sh} \simeq 10^8\,{\rm K}$. For both cases, the stellar mass-loss rate is set to $\dot{m}_{\rm w}=10^{-8}\,M_{\odot}{\rm yr^{-1}}$ and the wind velocity to $v_{\rm w}=100\,{\rm km\,s^{-1}}$. In Fig.~\ref{img_brems}, we plot the bremsstrahlung light curves for both layers. In case of the shocked wind, the luminosity increases because of the dominant density term -- the density in the shocked layer increases towards the pericentre. For the shocked ambient medium, the overall luminosity decreases because of the dominant size term, i.e. the size of the bow shock decreases. The bremsstrahlung signal from the shocked wind should dominate, but its luminosity is at least three orders of magnitude smaller than the quiescent luminosity of Sgr~A*, $L[2-10\,{\rm keV}]\sim 2\times 10^{33}\,{\rm erg\,s^{-1}}$ \citep{2017IAUS..322....1H}. The X-ray flux from the hypothetical PWN is expected to be stronger, at least at the level of the quiescent emission of Sgr~A*.

The low level of the X-ray bremsstrahlung emission for a stellar bow shock is not surprising. During the passage of S2 star through the pericentre in 2002, there was no significant increase in the emission of Sgr~A* detected \citep{2016MNRAS.456.1438Y}. It is worth monitoring the flaring rate during the next passage in early 2018, since some theoretical calculations predict the emission at the level of the quiescent state \citep{2016MNRAS.459.2420C}.    

\subsection{Br$\gamma$ emission model for a young star}
 
 One of the prominent features of the DSO are the recombination emission lines, mainly Br$\gamma$ line that is used for monitoring. So far and quite suprisingly, the luminosity of Br$\gamma$ line has remained constant within uncertainties \citep{Gillessen2013a, Gillessen2013b, Valencia2015}. It is not straightforward to explain the plateau of the light curve within the stellar model. In the framework of a photoevaporating and tidally disrupted protoplanetary disk, \citet{Murray-Clay2012} predict that the total Br$\gamma$ luminosity at the peribothron should be larger by a factor of five with respect to 2011 epoch. In case Br$\gamma$ recombination line originates due to the collisional ionization in the bow shock, the luminosity should also increase by a factor of a few \citep{Scoville2013,zajacek2016}. 
 
 One of the simplest solutions to the plateau of Br$\gamma$ line is to look for processes that are independent of the distance from Sgr~A* and of the velocity of the DSO. Circumstellar outflows and inflows that originate within the tidal radius of the DSO $r_{\rm t}$ (see Section~\ref{sec_compactness}) belong to such processes as was suggested by \citet{Valencia2015}. A likely process contributing to Br$\gamma$ emission in class I objects is the accretion of gas from the inner rim of an accretion disc that is truncated by a dipole magnetic field. The inferred large line width of Br$\gamma$ line of the order of $FWHM_{\rm Br\gamma} \approx 100\,{\rm km\,s^{-1}}$ \citep{Phifer2013,Valencia2015} is consistent with the magnetospheric accretion model \citep{1994ApJ...426..669H,1994AJ....108.1056E,1998ApJ...492..743M}, where the disc gas is channeled along magnetic field lines from the truncation radius $R_{\rm T}$ onto the stellar surface. The poloidal velocities are close to a free-fall velocity,
 
 \begin{equation}
   v_{\rm pol}=618\,\left(\frac{M_{\star}}{1\,M_{\odot}}\right)^{1/2}\,\left(\frac{R_{\star}}{1\,R_{\odot}}\right)^{-1/2}\,f_{\rm T}\,{\rm km\,s^{-1}}\,,
   \label{eq_poloidal_velocity}
\end{equation}  
where $f_{\rm T}=\sqrt{1-R_{\star}/R_{\rm T}}$ is the truncation factor, correcting the free-fall velocity for the finite truncation radius. Pre-main-sequence stars exhibit a considerable magnetic field with an intensity of $\sim 1\,{\rm kG}$, which truncates an accretion disc at several stellar radii,
\begin{equation}
  \frac{R_{\rm T}}{R_{\star}}\approx 6.5 B_{3}^{4/7}R_{2}^{5/7}\dot{M}_{-8}^{-2/7}M_{1}^{-1/7}\,,
  \label{eq_truncation_radius}
\end{equation}  
where the strength of the dipole magnetic field at the equator of the star $B_{3}$ is in ${\rm kG}$, the stellar radius $R_2$ is in units of $2R_{\odot}$, the accretion rate $\dot{M}_{-8}$ is in $10^{-8}\,M_{\odot}\,{\rm yr^{-1}}$, and the stellar mass $M_1$ is expressed in $1\,M_{\odot}$\footnote{The relation for the truncation radius, Eq.~\eqref{eq_truncation_radius}, gives an upper limit, since the ram gas pressure is larger in the disc geometry than in a spherical approximation that was used for this derivation}. Using an estimate in Eq.~\eqref{eq_truncation_radius}, the truncation factor in Eq.~\eqref{eq_poloidal_velocity} is $f_{\rm T}\approx 0.9$, which leads to $v_{\rm pol}\approx 560\,{\rm km\,s^{-1}}$.  
The line-of-sight velocity of the accreting gas that determines the Doppler broadening of an emerging line depends on the viewing angle of the star-disc system, which is itself variable for any structure bound to the star, e.g. a bow shock or a disc (see Appendix~\ref{appa1}). 

A stationary accretion can only proceed when the disc truncation radius is inside the corotation radius, which is defined as the distance from the star where the Keplerian angular velocity equals to the rotational angular velocity of the star,

\begin{equation}
  R_{\rm co}\approx 4.2\,M_1^{1/3}P_{1}^{2/3}\,R_{\odot}\,,
\end{equation}
where $P_1$ is the stellar rotational period in units of one day. 

\begin{figure}[tbh]
 \centering
 \includegraphics[width=0.5\textwidth]{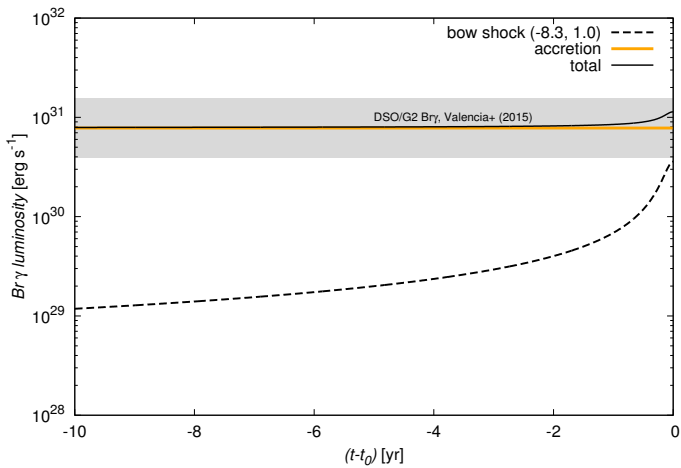}
 \caption{A temporal, synthetic evolution of the luminosity of Br$\gamma$ line for the bow shock contribution, the accretion contribution, and the total light curve. The wind parameters of the star were set to $\dot{m}_{\rm w}=5 \times 10^{-9}\,M_{\odot}\,{\rm yr^{-1}}$ and $v_{\rm w}=10\,{\rm km\,s^{-1}}$ (logarithms of these values are in the parentheses). The shaded region corresponds to the constant, observationally inferred luminosity of Br$\gamma$ line \citep{Valencia2015}.}
 \label{fig_brgamma}
\end{figure}

For pre-main-sequence stars, a correlation was found between the accretion luminosity $L_{\rm acc}$ and the luminosity of Br$\gamma$ emission. The correlation does not have to necessarily imply that Br$\gamma$ is a tracer of accretion \citep{2015MNRAS.452.2837M} or that the emission originates within a few stellar radii from the star. We can, however, use the $L_{\rm acc}$--$L_{\rm Br\gamma}$ correlation to show that a realistic accretion rate can reproduce the observed, approximately constant Br$\gamma$ luminosity of $L_{\rm Br\gamma} \approx (1-4) \times 10^{-3}\,L_{\odot}$ for the DSO \citep{Gillessen2013a, Gillessen2013b, Pfuhl2015,Valencia2015}. Applying the correlation inferred from a recent fit to several low-mass stars \citep{2014A&A...561A...2A}, 

\begin{equation}
  \log{(L_{\rm acc}/L_{\odot})}=1.16(0.07)\log{(L_{\rm Br\gamma}/L_{\odot})} + 3.60(0.38)\,,
  \label{eq_lacc}
\end{equation} 
we get the accretion luminosities $L_{\rm acc}\approx 1.3-6.6\,L_{\odot}$. Hence, the accretion luminosity may be in principle a significant fraction of the continuum emission of the DSO. 

Having calculated an order of magnitude of the accretion luminosity allows us to estimate the accretion rate $\dot{M}_{\rm acc}$, assuming a certain value for the truncation radius $R_{\rm T}\approx 5 R_{\star}$,
\begin{align}
  \dot{M}_{\rm acc} & \simeq  \frac{L_{\rm acc} R_{\star}}{G M_{\star}}\left(1-\frac{R_{\star}}{R_{\rm T}} \right)^{-1}\, \notag\\
                    & \approx  4.1 \times 10^{-8} \left(\frac{L_{\rm acc}}{L_{\odot}} \right) \left(\frac{R_{\star}}{R_{\odot}} \right) \left(\frac{M_{\star}}{M_{\odot}} \right)^{-1}\,M_{\odot}{\rm \,yr^{-1}}\,.
\end{align}
Given the constraints on the mass and the radius of the DSO based on the bolometric luminosity of $30\,L_{\odot}$ (see Fig.~\ref{fig_hr_dso}), $M_{\rm DSO} \lesssim 3\,M_{\odot}$ and $R_{\rm DSO}\lesssim 10\,R_{\odot}$, respectively, the upper limit of the accretion rate is $\dot{M}_{\rm acc}\leq 10^{-7}\,M_{\odot}{\rm yr^{-1}}$. The gas accretion in young stellar systems seems to be linked to stellar and circumstellar outflows with the rate of $\dot{M}_{\rm w}\approx 0.001-0.1\,\dot{M}_{\rm acc}$ \citep{2006ApJ...646..319E}.

Hence, the simple model of the accretion described above complements the previous components of the DSO model analysed in Section~\ref{sec_modelling}, see also Fig.~\ref{fig_rgb_image} for an illustration.   

The accretion of gas onto the star and associated outflows can in principle stabilise the light curve of Br$\gamma$ emission of the DSO, see Fig.~\ref{fig_brgamma}. While the emission associated with the bow shock is expected to increase towards the pericentre, the accretion luminosity may stay rather constant since accretion flows are located well inside the circumstellar tidal radius. For a certain set of parameters, the total luminosity can be constant within uncertainties. In Fig.~\ref{fig_brgamma}, the outflow rate of the star is set to $5\times 10^{-9}\,M_{\odot}{\rm yr^{-1}}$ and the terminal wind velocity is $10\,{\rm km\,s^{-1}}$.

\section{Summary}
\label{sec_summary}

The previous analysis of NIR observational data has shown that the NIR-excess source DSO/G2 is a peculiar object in the S-cluster, which was further supported by the detection of a high linear polarization degree of $30\%$ in the NIR $K_{\rm s}$ band. Taking into account the compact behaviour during the closest passage around the SMBH, we focused on possible compact, stellar models of the fast-moving source. It appears that the SED characteristics of the DSO are reproduced best by either a young, embedded stellar object with a non-spherical dusty envelope (consisting of a dusty envelope, bipolar cavities, and a bow shock) or, on the other hand, a compact remnant, specifically a pulsar wind nebula. In both cases, the objects are expected to move supersonically close to Sgr~A*.

Given the star-formation potential of the Galactic centre region, we first investigated the scenario of a pre-main-sequence star of class I that is still embedded in dense, optically thick dusty envelope. We found a composite stellar model surrounded by a non-spherical envelope (a star--dusty envelope--bipolar cavities--bow shock) that could reproduce the main characteristics of the continuum emission, including the linear polarization. In the framework of this model, bipolar cavities and the bow shock serve as sites for scattering stellar and dust photons. Furthermore, they increase the overall deviation of the source from the spherical symmetry. The $K_{\rm s}$ band continuum emission is dominated by scattered dust photons, which explains the significant polarized emission in this band. The polarization degree of $\sim 30\%$ is consistent with an optically thick dusty envelope and is comparable to other observed embedded young stellar objects in molecular cloud regions.

Moreover, we also looked at the neutron star model that could explain the SED as well as a high linear polarization degree -- more specifically, the DSO would exhibit similar characteristics as a young, Crab-like pulsar wind nebula. Although energetically it would be possible to reproduce the NIR flux densities and the spectral slope, given the non-detection of X-ray or radio counterparts makes this model speculative. On the other hand, our estimates demonstrated that even current NIR facilities could be used to search for candidates of young pulsar wind nebulae in the innermost parsec of the Galactic centre in the NIR domain. Such sources would actually manifest themselves as highly polarized, apparent NIR-excess sources, with flux densities comparable to the DSO. Some of the NIR-excess sources observed in the S cluster region could thus be suitable candidates for pulsar wind nebulae. The imaging with the current and future NIR instruments can thus complement standard radio searches for young neutron stars in the Galactic centre region, which will allow us to put more constraints on the neutron star population in the innermost parsec. 

Future monitoring of the DSO and other NIR-excess sources, mainly with the focus on the radiative properties (both continuum and line emission) and the dynamics, will further narrow down the scenarios to explain these intriguing sources in the central arcsecond of the Galactic centre.

\begin{acknowledgements} 
  We thank an anonymous referee for very useful suggestions that improved the paper. MZ is grateful for the hospitality of the Astronomical Institute of the Academy of Sciences of the Czech Republic where a part of this paper was written. MZ thanks all the participants of the Cologne-Prague-Kiel meeting 2016 (CPK16) for the discussion and the input. MZ and MP are members of the International Max Planck Research School for Astronomy and Astrophysics at the Universities of Bonn and Cologne. This project received the support of the Czech Science Foundation grant ``Albert Einstein Center for Gravitation and Astrophysics" (No 14-37086G). A part of the project was supported by the collaboration within SFB956--A2 (``Conditions for Star Formation in Nearby AGN and QSO Hosts") at the Universities of Cologne and Bonn.
\end{acknowledgements}


\appendix
\section{Viewing angle of the bow shock}
\begin{figure}[tbh]
  \centering
  \includegraphics[width=0.5\textwidth]{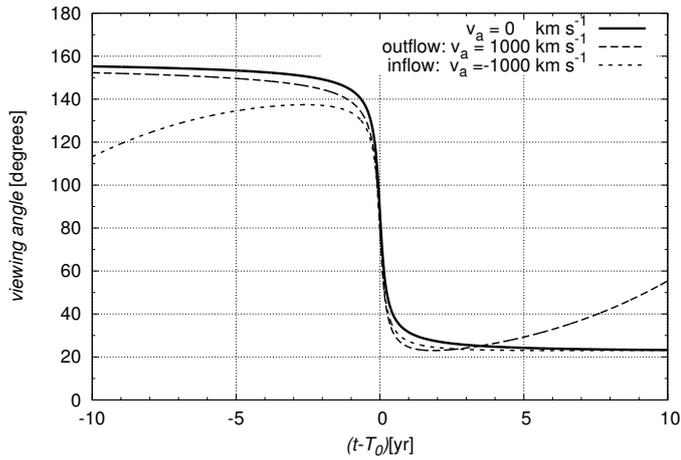}
  \caption{The viewing angle of the bow shock as a function of time for a negligible motion of the medium, an outflow of $1000\,{\rm km\,s^{-1}}$, and an inflow of $1000\,{\rm km\,s^{-1}}$ (see the key).}
  \label{fig_view_angle}
\end{figure}

We present the calculation of an expected angle between the symmetry axis of the bow shock structure and the line of sight as a function of time (with respect to the peribothron). In the calculation, we use the orbital elements as derived by \citet{Valencia2015}. We use the convention, in which $0$ degrees corresponds to the front view of the bow shock, $180$ degrees is the view from the tail portion, and $90$ degrees corresponds to the side view. We include three possible cases: a negligible motion of the medium, an outflow of $1000\,{\rm km\,s^{-1}}$, and an inflow of $1000\,{\rm km\,s^{-1}}$, see Fig.~\ref{fig_view_angle}.
\label{appa1}

\end{document}